%% file: main_full.tex
\documentclass[10pt,journal,compsoc]{IEEEtran}

\ifCLASSOPTIONcompsoc
  \usepackage[nocompress]{cite}
\else
  \usepackage{cite}
\fi

\ifCLASSINFOpdf
   \usepackage[pdftex]{graphicx}
   \graphicspath{{../pdf/}{../jpeg/}}
   \DeclareGraphicsExtensions{.pdf,.jpeg,.png}
\else
   \usepackage[dvips]{graphicx}
   \graphicspath{{../eps/}}
   \DeclareGraphicsExtensions{.eps}
\fi

\usepackage{caption}
\usepackage{graphicx}
\usepackage{booktabs}
\usepackage{colortbl}

\usepackage{array,tabularx}

\usepackage{subcaption}

\usepackage[english]{babel}
\usepackage[utf8x]{inputenc} 

\usepackage[super]{nth}
\usepackage[hyphens]{url}
\usepackage[hidelinks]{hyperref}
\hypersetup{breaklinks=true}
\graphicspath{{graphics/}}

\usepackage[font=small,labelfont=normalfont]{caption}
\usepackage[ruled,linesnumbered,vlined]{algorithm2e}
\usepackage{xcolor}

\SetCommentSty{mycommfont}

\begin{document}
\title{Hadoop Perfect File: A fast access container for small files with direct in disc metadata access}

\author{Jude~Tchaye-Kondi,Yanlong~Zhai,~\IEEEmembership{Member,~IEEE,}
		Kwei-Jay~Lin,~\IEEEmembership{Fellow,~IEEE,}
		Liehuang~Zhu,~\IEEEmembership{Member,~IEEE,}
		Wenjun~Tao,
		and Kai~Yang
\IEEEcompsocitemizethanks{\IEEEcompsocthanksitem J. Tchaye-Kondi, Y. Zhai, L. Zhu and W. Tao are  with School of Computer Science,
Beijing Institute of Technology, Beijing 100081, China.\protect\\
E-mail: tchaye59@yahoo.fr, ylzhai@bit.edu.cn
\IEEEcompsocthanksitem Prof. K.J. Lin is with Department of Electrical Engineering and Computer Science, University of California, Irvine 92697, CA US\protect\\
E-mail:  klin@uci.edu
\IEEEcompsocthanksitem K. Yang is with Science and Technology on Special, System Simulation Laboratory, Beijing Simulation Center, Beijing, China\protect\\
E-mail:  yangkai101@163.com}% <-this % stops an unwanted space
\thanks{Manuscript received April 19, 2005; revised August 26, 2015.}}

\IEEEtitleabstractindextext{%
\begin{abstract}

Storing and processing  massive small files is one of the major challenges for the Hadoop Distributed File System (HDFS). In order to provide fast data access, the NameNode (NN) in HDFS maintains the metadata of all files in its main-memory. Hadoop performs well with a small number of large files that require relatively little metadata in the NN's memory. But for a large number of small files, Hadoop has problems such as  NN memory overload caused by the huge metadata size of these small files.
We present a new type of archive file, Hadoop Perfect File (HPF), to solve HDFS's small files problem by merging small files into a large file on HDFS. Existing archive files offer limited functionality and have poor performance when accessing a file in the merged file due to the fact that during metadata lookup it is necessary to read and process  the entire index file(s). In contrast, HPF file can directly access the metadata of a particular file from its index file without having to process it entirely. 
The HPF index system uses two hash functions: file's metadata are distributed through index files by using a dynamic hash function and, for each  index file, we build an order preserving perfect hash function that preserves the position of each file's metadata in the index file. The HPF design will only read the part of the index file that contains the metadata of the searched file during its access. 
HPF file also supports the file appending functionality after its creation. 
Our experiments show that HPF can be more than 40\% faster file's access from the original HDFS.
If we don't consider the caching effect,
HPF's  file access is around 179\% faster than   MapFile and 11294\% faster than  HAR file. If we consider caching effect, HPF is around 35\% faster than  MapFile and 105\% faster than HAR file.

\end{abstract}

\begin{IEEEkeywords}
Hadoop, Small files, Distributed file system, Performance optimization, Archive files, Fast files access.
\end{IEEEkeywords}}

\maketitle

\IEEEdisplaynontitleabstractindextext

% \fi
\IEEEpeerreviewmaketitle

\begin{figure}[h]
  \centering
  \includegraphics[width=\linewidth]{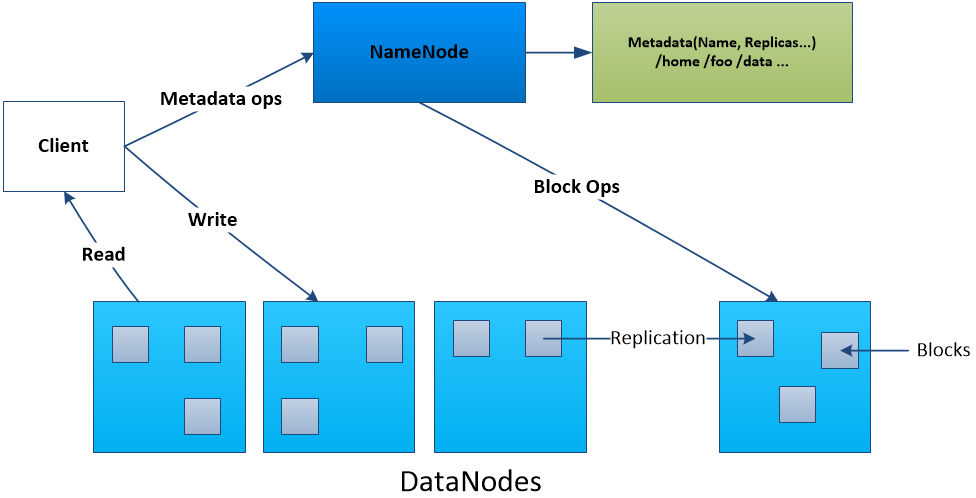}
  \caption{HDFS Architecture}
	\label{fig:hadoop_arch}
\end{figure}

\IEEEraisesectionheading{\section{Introduction}\label{sec:introduction}}
 \IEEEPARstart{T}{he} main purpose of Hadoop\cite{shvachko2010hadoop} is for efficient and swift storage and processing of big data. Hadoop's File System (HDFS)\cite{shvachko2010hadoop} uses a master-slave architecture (Figure \ref{fig:hadoop_arch}) based on GFS \cite{ghemawat2003google} to store and access data.
The entire HDFS is managed by a single server called the NameNode (NN) as the master, and  file contents stored on  DataNodes (DNs) as slaves. 
By default, each HDFS data block size is 128 MB but configurable according to the I/O performance that the client wants. 
To store a big file, Hadoop  splits it into many data blocks and stores them in different DNs. With Hadoop's built-in replication system, each
data block is replicated on several DNs (3 by default) to avoid data loss in case of a DN failure.

HDFS is very efficient when storing and processing large data files. But for a large number of small files, HDFS faces the  \textbf{small file problem}.
Applications such as social networks, e-commerce websites, digital libraries generate a large amount of data but in the form of small files. Many of these applications use data from healthcare, meteorology, satellite images, servers log files, etc.
For example, server’s applications generate many log files; depending on its configuration, an application can generate  a log file per hour or daily. Websites are often hosted on servers  in the cloud.  Regardless of the size of a website, log analysis can provide direct answers to problems encountered on websites. %Whenever a user or robot visits a website, a line or some line of logs will be written to a log file on the server. A visit can come from a visitor on mobile, tablet, PC or a robot like Google (Google Bot). 
The log analysis can identify the SEO traffic and observe the passage of Google's robots on a website as well as information on the website errors. Log analysis is useful for performing SEO audits, to debug optimization issues, to monitor the health of a website and its natural referencing.
Such data files are often small in size, from some KB to several MB, but very important for data analysis by data researchers. 

%, this situation pushes big companies to design their own solutions.
%When a file size is smaller than the block size this file is considered as small file and to store it Hadoop allocate a block with a size smaller than the default block size and equal to the file size.

%HDFS is very efficient when storing and processing large data files. But for large number of small files, HDFS faces the problem called \textbf{small file problem}.
Currently, there is no effective DFS that works  well for massive small  files.
Since  an NN usually maintains all its metadata in memory, massive small files generate much more metadata and could result in an {overload of NN's memory} which greatly affects its performance.
In addition to NN's memory overload, other problems caused by massive small files include:
\begin{enumerate}
	\item \textbf{Long storage time:} In our experiment, it could take up to 11 hours to upload 400,000 files of sizes ranging from 1 KB to 10 MB.
	\item \textbf{Bad processing performance:} A large number of small files means that MapReduce\cite{dean2008mapreduce} has to perform a large number of reads and writes on different nodes of a cluster, thus taking much more time than reading one big file.
	\item \textbf{NameNode  performance:} If multiple clients try to access many files at the same time, this will affect NN performance and also increase its memory overhead. In this situation, NN needs more time to process requests and to execute certain tasks.
\end{enumerate}

To overcome the problem of small files, Hadoop provides  HAR file, SequenceFile, and MapFile that can be used to reduce the NN memory load by combining small files into large, merged files.
The problem with such archive files is that as a side effect, they deteriorate the access performance of small files inside of large files due to the fact that the metadata that was kept in the NN memory for quick access are rebuilt in index file(s) and stored as normal files on HDFS.
Accessing a small file from an archive file without considering any caching effect may be done in three ways:
\begin{itemize}
	\item In the best case, the index file is read entirely to get the small file metadata (file position in the big file, file size, etc) and then the file content can be recovered from the merged file like with MapFile.
	\item In another case, it is required to read entirely many index files to get the small file's metadata before recovering the file content. This is the case of HAR file which uses two index files.
	\item In the worst case, if the archive file is not index based, to access a small file's content can require reading the merged file from the beginning to the end until the searched file is found.
\end{itemize}
In the index-based archive files, to retrieve a file’s metadata without the caching effect, it is necessary to read all index file(s) which leads to a surplus of I/Os operations and increase the file access time.
For large index files their reading and processing become very expensive in time, and greatly affects the throughput. 
To prevent reading index files at every file access, the archive files solutions usually maintain their index files' data in the client memory by using caching and prefetching techniques, although  the client memory may be very limited. 

In this paper, we present a new design of index-based archive files called Hadoop Perfect File (HPF). %which also consists of combining small files into large files and building index files before storing on HDFS.
The major innovation of our approach is that it provides direct and fast file’s metadata access within the index file even without the use of in memory caching techniques. Instead of reading all  index files and loading the metadata of all small files in memory before retrieving the metadata of the searched file, HPF can read only the part of the index file that contains the metadata of the searched file.
By using an order preserving minimal perfect hash function in its index system, HPF can calculate the offset and the limit of the index file to read the metadata of a file. After the offset and limit calculation, HPF seeks in the index file at the offset position to get the file's metadata.  Moreover, since seek to some random positions in a file is an operation that can take a long time when the file is very large, we avoid getting big index files by distributing the small files' metadata into several index files using an extendible hash function. Finally, 
in addition to providing fast access to files metadata, HPF also supports adding more files after its creation which is not the case of HAR files. And HPF does not force the client to sort the files before creating the archive file or adding files as does MapFile.   

The rest of the paper is structured as follows.
Section \ref{sec:RelatedWork} reviews related work. 
Section \ref{sec:MotivationForHPFDesign} shows the motivations of our design. 
Section \ref{sec:ProposedSolution} presents the design of HPF.
In Section \ref{sec:ImplementationOptimizations} we investigate some issues of our implementation. 
Section \ref{sec:ExperimentalEvaluation} evaluates the performance of our HPF implementation  against the native HDFS, HAR file, MapFile and analyzes experimental results.
Finally, the paper is  concluded in Section \ref{sec:ConclusionAndFutureWork}.

\section{Related Work}
\label{sec:RelatedWork}
In this section, we will discuss the solutions that have been proposed to deal with the problems of small files.
The researches on the problem of small files in DFS leads to three types of solutions.

\subsection{Combining small files into large files}
\label{sec:ArchiveFiles}

\begin{figure}[htbp]
	\begin{subfigure}{.5\linewidth}
		\centering
		\includegraphics[width=\linewidth]{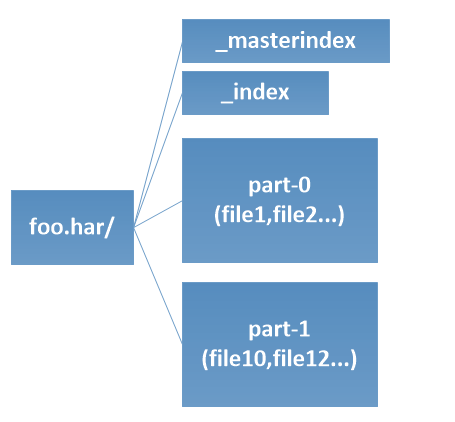}
		\caption{HAR file}
		\label{fig:har_lay}
	\end{subfigure}
	\begin{subfigure}{.5\linewidth}
		\centering
		\includegraphics[width=\linewidth]{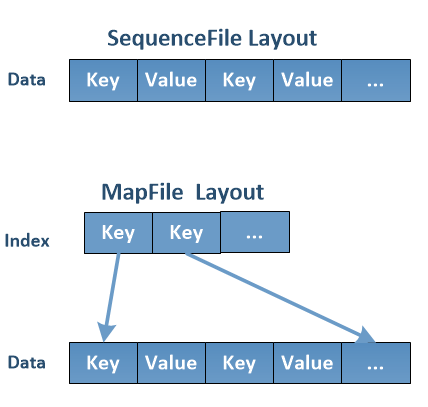}
		\caption{SequenceFile/MapFile}
		\label{fig:sequencefile}
	\end{subfigure}%
	\caption{ }
	\label{fig:sequencefile_har_lay}
\end{figure}

The first type of solution is to combine the small files into large files in order to reduce the amount of metadata needed for their storage.
HDFS come by default with some solutions like HAR file, SequenceFile, MapFile used to merge small files into large files.
HAR\cite{harSite} file is an archive file that keeps it files metadata through two index files: \_index and \_masterindex. As shown in Figure \ref{fig:har_lay} the small files' contents are stored in part-* files.
The weakness of HAR files is that they are immutable: Once created, we cannot modify its content anymore by adding or appending files inside, which is necessary when the small files are generated continuously. And also, HAR access performance is not very good.
The SequenceFile is proposed by Hadoop initially to solve binary log problem\cite{White15}. In this format, the data is recorded in the form of key-value pair (Figure \ref{fig:sequencefile}) where the key and the value are written in binary. 
%Although it was not originally created to hold small files, the SequenceFile proved to be a very good container for small files.
%SequenceFile supports block-level and record-level compression and file append functionality.
The well-known limitation of SequenceFile is that when searching for a file, it needs to traverse elements in the SequenceFile which has worst case complexity of O(n).
In order to overcome SequenceFile limitation, Hadoop proposes the MapFile, which is a sorted SequenceFile with an index to permit lookups by key as illustrated in Figure \ref{fig:sequencefile}.
Instead of storing every key in the index file, the MapFile index only stores every 128th key by default. So the complexity of accessing a file would be O(logn).
%The MapFile allow to quickly recover files in the data file, it firstly get from the index file which is small in size and loaded in memory, the file’s information before reading it from the data file.
The weakness of MapFile is that the client must sort files according to their keys before creating the MapFile
and also the MapFile does not support adding any file after its creation.%, the key of the file to be added must necessarily be in the lexicographical order.
%Due to Hadoop's default solutions to the small files problem limitation, researches have been carried out in order to find an optimal %solution. Some of these researches are presented in Table \ref{tab:other_solutions}.

Hadoop's default solutions present some limitations like the bad files access performance, the difficulty or impossibility of adding files after the creation of the archive. 
Numerous studies have attempted to solve the small files problem.
%Some researches have been carried out in order to find an optimal solution. 
In \cite{zheng2017method}, Tong Zheng et al. presented an approach that consists of storing the file's metadata in the HBase database after combining them into large files and using a prefetching mechanism by analyzing the access logs and putting the metadata of frequently accessed merge files in the client’s memory.
NHAR\cite{nhar} (New HAR) combines the small files into large files and distributes their metadata in a fixed number of index files by using a hash function.
%NHAR like HAR use MapReduce for their creation, but unlike HAR, NHAR supports the files appending functionality.
%Despite the efforts of NHAR to improve the performance of the HAR, the 
NHAR and the HAR still suffer from the slowness of their creations due to the fact that they need to upload all file to HDFS before creation.
Kyoungsoo Bok et al. proposed a distributed caching scheme to efficiently access small files in Hadoop distributed file system\cite{bok2017efficient}.
For improving the efficiency of storing and accessing small files on HDFS in BlueSky\cite{dong2010novel}   , one of the most prevalent eLearning resource sharing systems in China, Bo Dong et al. have designed a novel approach. In their approach, firstly, all correlated small files of a PPT courseware are merged into a larger file to reduce the metadata burden on NameNode. Secondly, a two-level prefetching mechanism is introduced to improve the efficiency of accessing small files.
%
%OMSS and TLB-MapFile have been proposed to improve the performance of Mapfiles.
OMSS(Optimized MapFile based Storage of Small files)\cite{sheoran2017optimized}% is proposed by Shalini Sheoran et al. and 
 is a new algorithm which merges the small files into a large file based on the worst fit strategy. The strategy helps in reducing internal fragmentation in data blocks.
%, which in turn leads to fewer data blocks consumed for the same number of small files. 
Less number of data blocks mean fewer memory overheads at NN and hence increased the efficiency of data processing. 
TLB-MapFile is proposed in \cite{meng2016novel} and is an access optimization approach for HDFS small file based on MapFile. 
TLB-MapFile% merges massive small files into large files by MapFile mechanism to reduce NameNode memory consumption and 
adds fast table structure(TLB) in DataNode to improve retrieval efficiency of small files. 
%First, according to MapFile mechanism, small files are merged into large files and stored in HDFS. 
%Second, 
The access frequency and the ordered queue of small files (per unit time) can be obtained through the system logs in HDFS, and the mapping information between block and small files are stored in the TLB table with regularly being updated. 
Since OMSS and TLB-MapFile are based on MapFile, they require the files keys to be in lexicographic order and therefore not optimize for random files add and accesses.
Some proposed solutions are in the direction of modifying HDFS by adding hardware to speed up the processing of small files or by making HDFS automatically combine small files before storage.
%
%This is the case of Jian-feng 
Peng et al. 
%who in order to improve the efficiency of storing and accessing the small files on HDFS, 
proposed a Small Hadoop Distributed File System (SHDFS)\cite{peng2018hadoop}, which is based on original HDFS
%. Compared to original HDFS, they add two novel modules in their proposed SHDFS: 
but added a merging module and a caching module.
 In the merging module, they proposed a correlated files model, which is used to find out the correlated files by user-based collaborative filtering and then merge correlated files into a single large file to reduce the total number of files. In the caching module, they use Log-linear model to dig out some hot-spot data that user frequently accesses to, and then design a special memory subsystem to cache these hot-spot data in order to speed up access performance.% to hot-spot data.
%
%Yonghua Huo and Zhihao Wang 
Hou et al. in their proposed solution use additional hardware named SFS\cite{huo2016sfs} (Small File Server) between users and HDFS to solve the small file problem. Their approach includes a file merging algorithm based on temporal continuity, an index structure to retrieve small files and a prefetching mechanism to improve the performance of file reading and writing.
%These approaches suffer from the fact that they modify HDFS.
Among the archive-based solutions, some before merging the files classify them into categories or rely on the distribution of files according to some criteria that help to optimize the storage or access performance like 
LHF\cite{EasyChair:773}, DQSF\cite{jing2018optimized}, Hui He et al.\cite{he2016optimization}, Xun Cai et al.\cite{cai2018optimization}.
In case the proposed solution is built on top HDFS, it is easy to migrate to the latest version of HDFS. But when the solution modifies HDFS, its maintenance and update becomes difficult and expensive for companies. Which is not desirable.

\begin{table*}
\centering
\setlength{\extrarowheight}{0pt}
\addtolength{\extrarowheight}{\aboverulesep}
\addtolength{\extrarowheight}{\belowrulesep}
\setlength{\aboverulesep}{0pt}
\setlength{\belowrulesep}{0pt}

\resizebox{\linewidth}{!}{%
\begin{tabular}{lcccccccc} 
\toprule
\rowcolor[rgb]{0.753,0.753,0.753} Paper Name / Feature                                                                                                                      & Type                 & \begin{tabular}[c]{@{}>{\cellcolor[rgb]{0.753,0.753,0.753}}c@{}}Master\\Memory\\usage\end{tabular} & \begin{tabular}[c]{@{}>{\cellcolor[rgb]{0.753,0.753,0.753}}c@{}}Support\\Append\end{tabular} & \begin{tabular}[c]{@{}>{\cellcolor[rgb]{0.753,0.753,0.753}}c@{}}Modify\\HDFS\end{tabular} & \begin{tabular}[c]{@{}>{\cellcolor[rgb]{0.753,0.753,0.753}}c@{}}Use\\extra\\System\end{tabular} & \begin{tabular}[c]{@{}>{\cellcolor[rgb]{0.753,0.753,0.753}}c@{}}HDFS\\pre-upload\\required\end{tabular} & \begin{tabular}[c]{@{}>{\cellcolor[rgb]{0.753,0.753,0.753}}c@{}}Creation\\Overhead\end{tabular} & \begin{tabular}[c]{@{}>{\cellcolor[rgb]{0.753,0.753,0.753}}c@{}}Reading\\Efficiency\\(Complexity)~~\end{tabular}  \\ 
\hline
HDFS                                                                                                                                                                        & DFS                  & Very High                                                                                          & Yes                                                                                          & -                                                                                         & -                                                                                               & Yes                                                                                                     & Very High                                                                                       & High                                                                                                               \\ 
\hline
HAR                                                                                                                                                                         & Archive\&Index Based & Low                                                                                                & No                                                                                           & No                                                                                        & No                                                                                              & Yes                                                                                                     & Very High                                                                                       & Low                                                                                                                \\ 
\hline
MapFile                                                                                                                                                                     & Archive\&Index Based & Very Low                                                                                           & For special keys                                                                             & No                                                                                        & No                                                                                              & No                                                                                                      & Moderate                                                                                        & High(O(logn))                                                                                                      \\ 
\hline
SequenceFile                                                                                                                                                                & Archive Based        & Very Low                                                                                           & Yes                                                                                          & No                                                                                        & No                                                                                              & No                                                                                                      & Low                                                                                             & Low(O(n))                                                                                                          \\ 
\hline
BlueSky\cite{dong2010novel}                                                                                                                                                      & Archive\&Index Based & Low                                                                                                & Yes                                                                                          & No                                                                                        & No                                                                                              & No                                                                                                      & High                                                                                            & High                                                                                                               \\ 
\hline
T. Zheng et al\cite{zheng2017method}                                                                                                                                             & Archive\&HBase Based & Low                                                                                                & Yes                                                                                          & Yes                                                                                       & Yes                                                                                             & No                                                                                                      & High                                                                                            & High                                                                                                               \\ 
\hline
NHAR\cite{nhar}                                                                                                                                                                  & Archive\&Index Based & Low                                                                                                & Yes                                                                                          & No                                                                                        & No                                                                                              & Yes                                                                                                     & High                                                                                            & High                                                                                                               \\ 
\hline
\begin{tabular}[c]{@{}l@{}}OMSS\cite{sheoran2017optimized},\\TLB-MapFile\cite{meng2016novel}\end{tabular}                                                                             & MapFile Based        & Very Low                                                                                           & For special keys                                                                             & No                                                                                        & No                                                                                              & No                                                                                                      & Moderate                                                                                        & High                                                                                                               \\ 
\hline
SHDFS\cite{peng2018hadoop}                                                                                                                                                       & Archive\&Index Based & Low                                                                                                & Yes                                                                                          & Yes                                                                                       & Yes                                                                                             & No                                                                                                      & High                                                                                            & High                                                                                                               \\ 
\hline
SFS\cite{huo2016sfs}                                                                                                                                                             & Archive\&Index Based & Low                                                                                                & Yes                                                                                          & Yes                                                                                       & Yes                                                                                             & No                                                                                                      & High                                                                                            & High                                                                                                               \\ 
\hline
LHF\cite{EasyChair:773}                                                                                                                                                          & Archive\&Index Based & Low                                                                                                & Yes                                                                                          & No                                                                                        & No                                                                                              & No                                                                                                      & Moderate                                                                                        & High                                                                                                               \\ 
\hline
\begin{tabular}[c]{@{}l@{}}DQSF\cite{jing2018optimized}, He\cite{he2016optimization},\\Cai\cite{cai2018optimization},Kyoungsoo\cite{bok2017efficient}\end{tabular} & Archive\&Index Based   & Low                                                                                                & Yes                                                                                          & No                                                                                        & No                                                                                              & No                                                                                                      & High                                                                                            & High                                                                                                               \\ 
\hline
HPF           
                                                                                                                                                              & Archive\&Index Based   & Low                                                                                                & Yes                                                                                          & No                                                                                        & No                                                                                              & No                                                                                                      & Moderate                                                                                        & Very High(O(1))                                                                                                   \\
\bottomrule
\end{tabular}
}
\caption{Comparison of solutions to small files problem}
\label{tab:othersolutions}
\end{table*}

\subsection{Specialized DFS for small files}
\label{sec:SpecializedDFSForSmallFiles}

The second type is to build DFS specialized only in the processing of small files like TFS\cite{tfs} used by Taobao, Haystack\cite{beaver2010finding} used by Facebook, Cassandra\cite{lakshman2010cassandra} used by Twitter.
Faced with the problem of massive small files, Facebook has to process about more than a million images per second. To ensure a good user experience, Facebook set up Haystack architecture. In this architecture, users’ pictures are combined in big files and the pictures' metadata are used to build the index file. The Haystack maintains all index files data in main-memory in order to minimize the number of disk operations to the only one necessary for reading the file content.
Taobao is one of China’s largest online marketplace also has to deal with the problem of small files.
Taobao generates about 28.6 billion photos with average size of 17.45 KB \cite{fu2015performance}. To provide high availability, high reliability and high performance, Taobao creates TFS (Taobao File System) \cite{tfs} , a distributed file system designed for small files less than 1MB in size. TFS is based on IFLATLFS\cite{fu2015performance} a Flat Lightweight File System and is similar to GFS. Unlike other file system IFLATLFS aims is to reduce the size of metadata needed to manage files to a very small size in order to maintain them all in memory. 
%For example, when Ext4 is used, 92.8 GB of metadata is needed to store 10 TB data with the average file size of 16 KB but with IFLATLFS 10 TB data will only need about 7.5 GB of metadata when the average file size is 16 KB \cite{fu2015performance}.

\subsection{Improving processing framework}
\label{sec:ImprovingAccessingAndProcessingEfficiencyOfSmallFiles}

The third type of solution consist only in improving the accessing and processing framework for small files in DFSs or traditional file systems.
Priyanka et al. have designed a CombineFileInputFormat to improve the performance of processing small files using MapReduce framework\cite{phakade2014innovative}. Normally, a map task takes as input a split which is a block of data at a time.
In the case of small files, as the file size is smaller than the size of the block, the map task receives a small amount of input data.
Their approach consists in combining several small files into big splits before providing them as input to the map task.
This approach has then been improved by Chang Choi et al. in \cite{choi2018improved} by integrating the CombineFileInputFormat and the reuse feature of the Java Virtual Machine (JVM). 
%Their method reduces the number of created mappers by processing large numbers of files that are combined by a single split using CombineFileInputFormat. When running MapReduce, a JVM is created for each map task, to improve MapReduce processing performance, their method reduces JVM creation time by reusing a single JVM to run multiple mappers.
%The work of Priyanka Phakade et al. and Chang Choi et al leads to improving the processing of small files by MapReduce but completely ignores NN memory overflow caused by the small files' metadata.
%
%The traditional Unix file system (ext2, ext3, ext4,etc) uses the inode data structure to store file and folder information. Whenever it is needed to access a file on Unix, the file’s inode information are read first. 
%Inode contain information like file type (regular file, directory.), file’s Permissions: (read, write, execute), link count, user ID (the owner), group ID, size of file, time stamps (access time, modification time,etc), Attributes: (immutable,etc), access control list, Link to location of file,etc.
There are some researches have attempted to modify the structure of the OS's file system to improve the access performance of small files\cite{niazi2018size}\cite{kim2017improving}.
%
%Inode contains too much information that is useless in some situations and reading these large amounts of information significantly slows down the files' access. 
%\cite{niazi2018size}\cite{kim2017improving} modify the inode structure by removing its useless information in order to minimize the IOs of its reading and writing.
%
\cite{niazi2018size} designed the stuffed inode for small files that embeds the content of small files in the inodes’ metadata in a variant of HDFS with distributed metadata called HopsFS\cite{niazi2017hopsfs}.   \cite{kim2017improving} modified both the in-memory and on-disk inode structure of the existing filesystem and were able to dramatically reduce the number of storage and access I/O operations.

Together these studies provide important insights into the massive small files problem, but none of them seems to meet all the criteria necessary to become the only solution for big data storage system with massive small files and also big files.
%
%
%In our above discussion, we presented three types of solutions against the problem of small files and we compare the characteristics of some in Table \ref{tab:othersolutions}.
The solutions that combine small files into large files effectively reduce the memory consumption of the NN, but most of they deteriorate heavily the files access performance because they require an overhead of IO operations to get metadata and file content.
Specialized DFS for small files does not work very well or does not support large files.
The solutions that only improve the small file processing framework or underlining file system bring no advance to the overload of NN memory.
As summarized in Table \ref{tab:othersolutions}, the detailed comparison of HPF with some existing solutions shows that our HPF is superior in all aspects.
In our work, we focused on putting in place an optimal storage solution which reduces the load on the NN memory and offers great access performance. The HPF is built on top of HDFS and require no modification of HDFS. At the mean time, HPF supports file append functionality with little cost. This makes HDFS very efficient with managing small files as well as large files.

\section{Motivation For hpf Design}
\label{sec:MotivationForHPFDesign}
 File access in HDFS and archive files is done in two steps.
The first set consists in to get the file’s metadata and the second step consist in to restore the contents of the file.
In HDFS files metadata are kept in memory of the NN whereas the archive files maintain their metadata in the index files.
HDFS provides a High access efficiency to DNs data by keeping their metadata in NN's main-memory. Each folder, file, and block generate metadata; in general, the metadata of a file consumes about 250 bytes of main memory. For each block with default 3 replicas, its metadata can consume about 368 bytes. With 24 million files in HDFS, NameNode will require 16 GB of main memory for storing metadata\cite{shvachko2007name}.
For each file, the NN keeps information about the file and information about each block of the file.

\subsection{Normal file access from HDFS}
\label{sec:FileAccessFromHDFS}

For a file stored directly in HDFS, read the contents of this file, this file is done as follows \cite{fu2015performance}:
\begin{itemize}
	\item $T_{1}$: The client sends a request containing the file path to the NN
	\item $T_{2}$: The NN looks for the file's metadata located in its main-memory
	\item $T_{3}$: The NN returns a response to the client with the file's metadata. The metadata contains the file's blocks and their locations (the DNs on which the client must read blocks contents)
	\item $T_{4}$: The client sends a request to the DN
	\item $T_{5}$: The DN reads the file from its storage space
	\item $T_{6}$: The DN returns a result containing the block content to the client
\end{itemize}
\begin{equation}
	T=T_{1}+T_{2}+T_{3}+T_{4}+T_{5}+T_{6}
\label{equ:readtime}
\end{equation}
It needs 4 network operations $ (T_{1}, T_{3}, T_{4}, T_{6}) $, 1 disk operation$ (T_{5}) $ and 1 in memory operation$ (T_{2}) $. In total, 6 operations are required and The total time needed for reading a file is calculated by using Equation \ref{equ:readtime}.\newline
The time consumed by network operations depend on the quality of the network, $T_{2}$ is negligible since the metadata are located in memory of the NN their lookup is very fast.
$T_{5}$ takes longer since the file is stored on DN's discs, reading it is very slow compare to reading data from memory.
To summarize, reading a file from HDFS is done in two steps. Firstly, we get the file information from the NN, then once its information is acquired, the client can finally read the content of the file from the DNs. Getting file information from the NN is a very fast operation since the NN keeps the metadata in memory unlike the reading of the file content form the DN where the file's content is written in the disc.

\subsection{File access from Archive file without caching}
\label{sec:FileAccessFromArchiveFileWithoutConsideringTheCachingEffect}

Index-based archive files merge small files and build one or many of index files to store their metadata. In order to get the content of  a small file in the big merged file, we must first read the file's metadata from these index files. 
For example, with the MapFile which consists of two files (data, index), a client must read the metadata of the small file he is looking for from the index file before reading the content of the file from the data file.
In the first step, read the index file will require 6 operations as discussed in subsection \ref{sec:FileAccessFromHDFS} and in the second step, recovering the file's content will also require 6 operations. A total of 12 operations including 8 network operations and 2 disk operations are required to access a file in MapFile. 
The performance degrades quickly the more index file levels the archive file has.
The HAR file offers bad access performance because it uses two index levels. The client must process the \_masterindex file and after the \_index file before access the part file to read the content of a file.
It is obvious that to improve the access performance of small files we have to minimize the number of disk operation to one, the only one needed to read the file content. That's why some archives files maintain their index files data in memory.

\subsection{File access from archive file with caching}
\label{sec:FileAccessFromArchiveFileWithTheCachingEffect}

In order to improve the file access performance within the archive files by avoiding the additional I/Os operations generate by index files, most of the archive files systems prefer maintain their index files information in memory.
When a client accesses a file from an HAR file for the first time, HAR file system will check if the HAR file's metadata is loaded in the client memory, if not, HAR file system will read the \_masterindex and \_index files and maintains their content in client's memory. During the next access HAR file system will not read the index files again and will get the file's metadata from the client memory, by default, HAR file system prefetch using the LRU\cite{dan1990approximate} algorithm the metadata of 10 HAR files in client’s memory. 
MapFile also maintains its metadata in client memory to improve its access performance, when a client firstly accesses a file in a MapFile, the MapFile loads its index file content into the client's memory. 
Keeping the metadata in the client memory poses no problem when the client has enough, but this can become a problem when the memory of the client is limited and that client have to access lot of archive files at the same time.

HAR file and MapFile seem to move NN's memory burden to client's memory to optimize their access performance, our solution does not need to maintain files metadata in client’s memory, we operate on DataNodes memory by using the Centralized Cache Management system of HDFS\cite{CentralizedCacheManagement}.
Our approach unloads the load generated by small files metadata in NN’s memory and uses less of the client memory which makes it very convenient for the client with a low memory capacity.

\section{Hadoop Perfect File}
\label{sec:ProposedSolution}
HPF is an index-based archive file and is built with the goal of very fast metadata lookup for all small files. It is also designed to support new file appending functionality. The most important contribution of HPF is that its index system is built to make the metadata lookup time from index files so small that it is almost negligible, so as to greatly increase the processing and accessing performance of small files.
\begin{figure}[h]
	\centering
		\includegraphics[width=\linewidth]{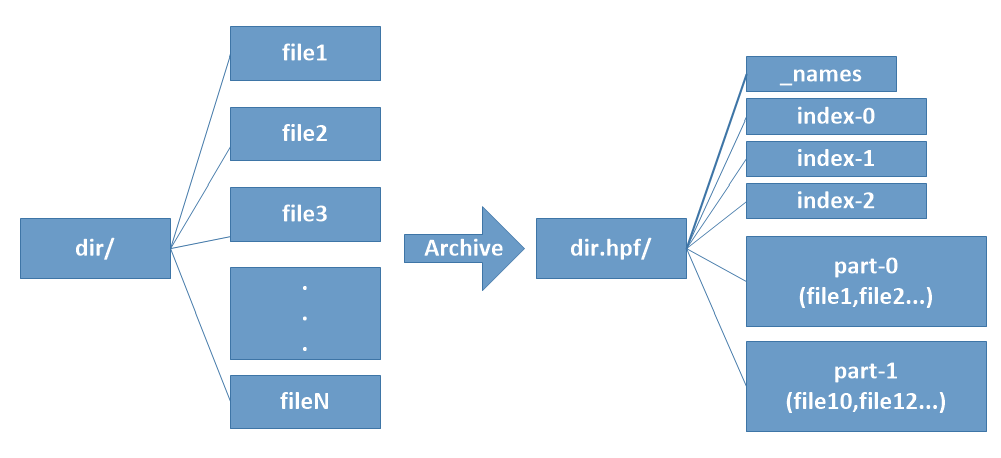}
	\caption{HPF folder}
	\label{fig:hpf_folder}
\end{figure}

\subsection{HPF Structure}
\label{sec:HPFStrcuture}

In our design, an HPF file is a folder (Figure \ref{fig:hpf_folder}) that contains many index-* files (index-0, index-1, etc.), some file collections called part-* files (part-0, part-1, etc.), and a \_names file. The index-* files contains  small files' metadata. The part-* files have the actual contents of the small files.  The \_names file  holds the names of all small files that have been appended to one of the part-* files,  so that users can find the names of all small files included in the HPF file.

\begin{figure}[h]
	\centering
		\includegraphics[width=\linewidth]{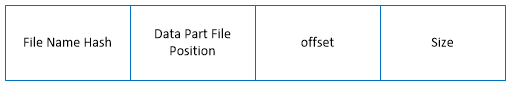}
	\caption{Small file metadata}
	\label{fig:bucket_entry}
\end{figure}

When creating an HPF file, we combine small files into large part-* files. To easily find the contents of  small files from the part-* files,   whenever a small file is appended to a part file, we create its metadata as in Figure \ref{fig:bucket_entry}. %We insert the metadata into the corresponding bucket using an extensible hash function.
A file's metadata has the following information:
\begin{itemize}
	\item \textbf{File name hash:} This is a unique integer that is generated by submitting the small file name to a hash function. This hash value is unique for each file and identifies a small file inside the archive file.
	\item \textbf{Data part file position:} This value is also an integer that represents which part file contains the small file. For example, position 0 means the small file is in the file part-0.
	\item \textbf{Offset:} This is the actual offset to read the small file from the specific part-* file.
	\item \textbf{Size:} This is the size of the small file.
\end{itemize}

One issue with HAR is that it uses two-level index which deteriorates its performance during files access. In our work, 
HPF uses only one level index but divides the information into many index files. As shown earlier, a small file's metadata  contains the  minimum information required to find the file's content form one of the part-* files. Each small file's metadata will be saved in one of the index-* files; a hash function will be used to decide which index-* file is used. Moreover, we use another hash function to quickly find the exact location of a small file's metadata  in that index-* file. In other words, the HPF index system uses two hash functions to identify a small file's metadata location. The first one is the extensible hash function(EHF)\cite{fagin1979extendible}\cite{zhang2009extendible} to decide which index-* file has the small file’s metadata. The second hash function is an order preserving minimal perfect hash function(MMPHF)\cite{fox1989order}, used to decide where in the index file to find a file’s metadata.

%HPF does not force the client to sort all small files during the merging process like MapFile, and it does not also force the client to upload all small files to HDFS like in HAR.

\begin{figure}[h]
	\centering
		\includegraphics[width=\linewidth]{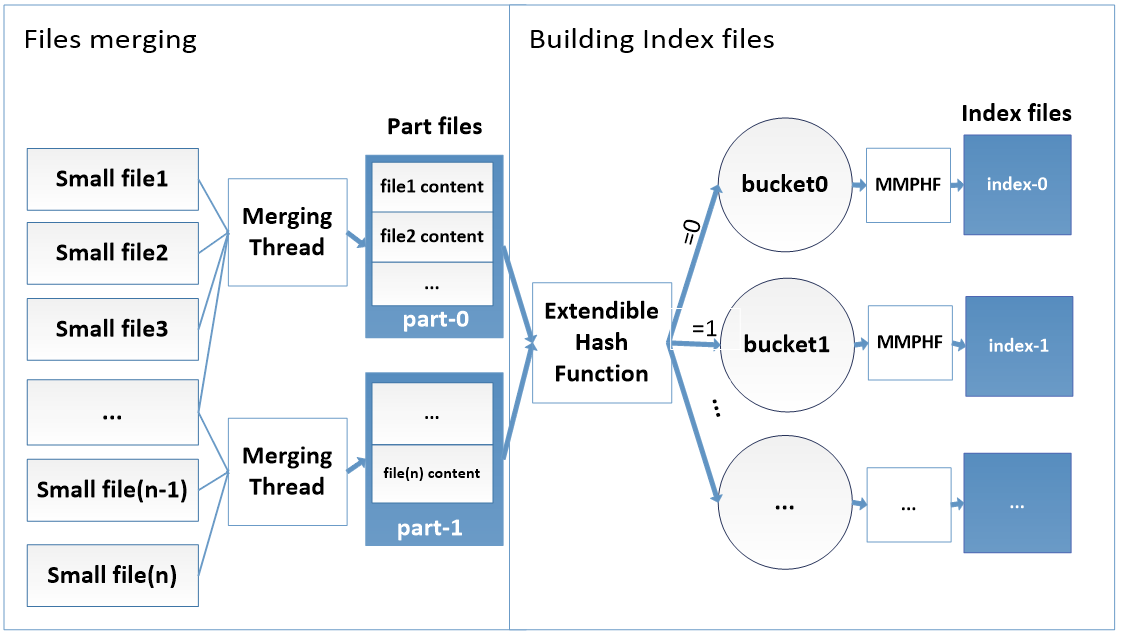}
	\caption{HPF Creation Flow}
	\label{fig:hpf_creation}
\end{figure}

The creation process of HPF file (Figure \ref{fig:hpf_creation}) can be summarized in 2 steps: (1) Merge small files into large part-* files, (2) Use the metadata of these small files to build index-* files.
In the first step which combines small files into large part-* files, we improve the NN's performance by reducing its memory burden.
In the second step, we create the index-* files that can be looked up quikcly to retrieve small files  from part-* files. Our index-* files design is very unique and truly efficient, which can be claimed as our major innovation in this work.
These two steps are discussed in the following two subsections.

\subsection{File merging process}
\label{sec:FilesMergeStrategy}

The merging step is illustrated by the files merging block of Figure \ref{fig:hpf_creation}.
The reason why HPF uses multiple part-* files instead of one is to support the small files merging in parallel for different part-* files.
Each merging thread merges small files to  one data part-* file. 
The number of parallel threads used during files merging is two by default; so we firstly create two HDFS data part-* files (\textbf{part-0} and \textbf{part-1}). 
We also create a temporary index file named \textbf{\_temporaryIndex} which will be used for recovery in case of failures and the \textbf{\_names} file which will contain all small files names.

For each small file in a merging thread, the content of the file is loaded into the client memory since it is faster when the file is at the client side than on HDFS. 
%If the file is on HDFS, loading its content in client memory will require some network I/Os and will take more time.
Once the small file content is in client memory, if the compression is enabled, we compress the content and then append the compressed data to the data part file; if not, we directly append the content to the part file. When appending to the part-* file finishes, we create the small file's metadata and insert it into HPF index system.
The data part files containing the contents of the small files are numbered like part-0, part-1, etc.

Initially, the HPF file has only two data part files: part-0 and part-1. If the client defined a maximum capacity for each part-* file, every time a file's content is to be appended to a part-(i) file, we check the size of the part-(i) file to see if it is larger than or equal to the maximum size defined for each part files. If it is the case, we create a new part file to append all future small files.

\subsection{Index files building process}
\label{sec:IndexSystemStrategy}

The metadata information is designed to have a fixed size for each small file, as shown in Table \ref{tab:IndexItemFieldsInformation}, taking 24 bytes in an index file. %To have this fixed size each field is represented by using primitive types: long (8 bytes), int (4 bytes). 
One issue is that file names do not have a fixed number of characters. To ensure the fixed metadata size, we have replaced each file name with a hash value which is an integer of a fixed size. %For any other field whose size may vary, we use an alternative fixed size primitive data types to represent it.
\begin{table}[h]
	\centering
		\begin{tabular}{|l|l|l|}
				\hline
				Field & Type & Size(bytes) \\ 
				\hline
				File Name Hash & \multicolumn{1}{c|}{long} & \multicolumn{1}{c|}{8} \\ 
				\hline
				Data Part File Position & \multicolumn{1}{c|}{int} & \multicolumn{1}{c|}{4} \\ 
				\hline
				offset & \multicolumn{1}{c|}{long} & \multicolumn{1}{c|}{8} \\ 
				\hline
				Size & \multicolumn{1}{c|}{int} & \multicolumn{1}{c|}{4} \\ 
				\hline
				\multicolumn{2}{|c|}{Total Size} & \multicolumn{1}{c|}{24} \\ 
				\hline
		\end{tabular}
	\caption{Metadata Fields Information}
	\label{tab:IndexItemFieldsInformation}
\end{table}

Given a small file’s metadata, we can just position the index file reader at the start offset and read from there to the end offset.
The operation that moves the reader to the designated offset of a file is the seek operation which, however, is expensive for random seeks between different blocks of the same file, since it takes time to establish a new connection with the block's DN.
To avoid the performance degradation due to random seek operations, we can limit the size of each index file so that it fits in only one HDFS block. 
%The maximum number of metadata that an index file can support is given by Equation~\ref{equ:index_maxRec}.
%\begin{equation}
%		Max\ metadata = \frac{Index\ file's\ Block\ Size}{File's\ metadata\ Size}
%\label{equ:index_maxRec}
%\end{equation}
If the block size of an index file is 128MB and a file's metadata occupies 24 bytes, the maximum number of  metadata can be stored in an index file will be $128*1024*1024/24 = 5,592,405$, which is large enough for practical purposes.

To limit the size of HPF’s index files an extensible hash function is used to dynamically distribute the metadata of files into multiple buckets saved as the index files on HDFS. The extendible hashing has been specially selected in our design for the capability of splitting an overflowed bucket easily.
The construction of  index files is done in two phases.
The first phase is performed simultaneously with the process of merging small files contents to large files. As shown in the index building block of Figure~\ref{fig:hpf_creation}. This phase consists of building buckets in the client memory by using the extensible hash table mechanism. 
In this phase, every time a file is appended to a part-* file, we append the file metadata to the \_temporaryIndex file,  the file name to the names file, and then we insert its metadata into the corresponding bucket of the extendible hash table.
In the second phase, for each bucket, we sort its records, construct an order preserving minimal perfect hash function, and write each bucket records in its corresponding index file.
The detailed design is presented in the following subsections.

\subsubsection{Inserting metadata in a bucket}

\begin{figure}[h]
	\centering
		\includegraphics[width=\linewidth]{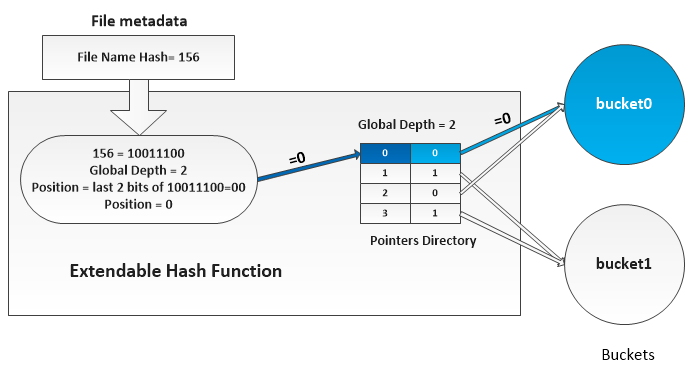}
	\caption{Example of inserting a file information into a bucket}
	\label{fig:ehf_insert}
\end{figure}

An Extendible Hash Table (EHT)\cite{fagin1979extendible}\cite{zhang2009extendible} is a dynamic hashing technique. As defined in~\cite{peb_eh}: 
\begin{quote} An EHT is a hash table in which the hash function is the last few bits of the key and the table refers to buckets. Table entries with the same final bits may use the same bucket. If a bucket overflows, it splits, and if only one entry referred to it, the table doubles in size. If a bucket is emptied by deletion, entries using it are changed to refer to an adjoining bucket, and the table may be halved.
\end{quote}
The number of last few bits that the EHT considers from a key to determine in which bucket to insert that key is called the global depth.  More information about the extendible hashing can be found in~\cite{Fagin:1979:EHF:320083.320092}.
When a new bucket is created, a new index file is created and associated with the bucket. To decide in which bucket to insert the metadata of a small file, the EHT looks at the last global depth bits of the hash value of the file name and considers the bucket pointed by its directory at the corresponding position (Figure~\ref{fig:ehf_insert}).
We serialize and store the EHT as an extended attribute\cite{ExtendedAttributes} of HPF file.
During HPF file creation, only one bucket is created and when the bucket reaches its maximum capacity (HDFS block size) another is created by calling the split operation.

\begin{figure}
	\centering
		\includegraphics[width=\linewidth]{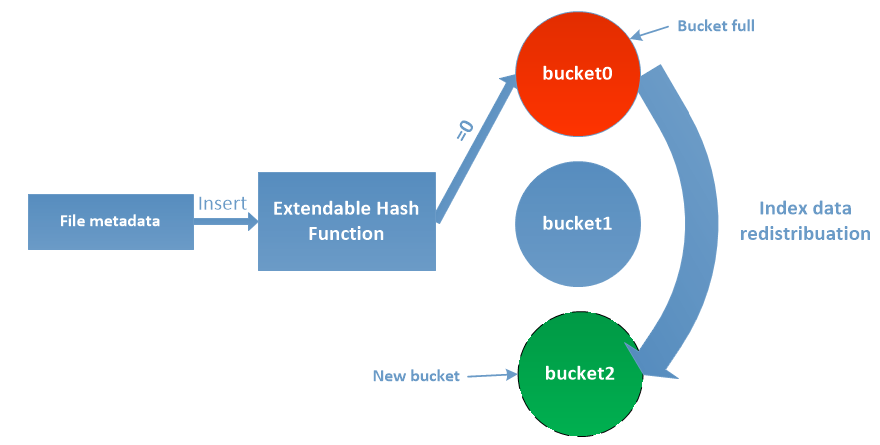}
	\caption{Index file bucket split}
	\label{fig:ehf_split}
\end{figure}

The split operation is an extendible hashing operation that is used to dynamically increase the number of buckets (index files) while providing direct access to the bucket during lookup.
%Each bucket has a maximum capacity, i.e. a maximum number of metadata that it can contain.
%After several insertions a bucket may reach its maximum capacity, in this case, we will split this bucket.
The bucket split process shown in Figure~\ref{fig:ehf_split} and is composed of two steps. The first is to create a new bucket, create and associate to this new bucket a new index file. The second step is to recalculate the positions of all records currently in the bucket having reached its maximum capacity and move those whose position has changed into the new bucket, this operation is called redistribution.
  
%After merging all small files contents with the part-* files, distributing their metadata into buckets and appending these metadata to the \_temporaryIndex file, we continue with the last step of the HPF file creation which consist in building the order preserving minimal perfect hash functions (OPMPHF) for each bucket. The role of this function is to return the position of a file's metadata from the index file when it is submitted to a file name hash value. We will discuss this function in subsection \ref{sec:IndexFilesOPMPHF}.

\subsubsection{Bucket monotone minimal perfect hash functions}
\label{sec:IndexFilesOPMPHF}

\begin{figure}[h]
	\centering
		\includegraphics[width=\linewidth]{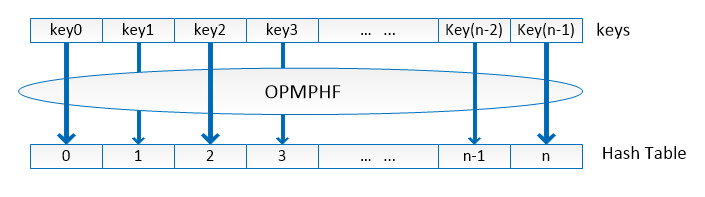}
	\caption{OPMHF}
	\label{fig:opmhf_lay}
\end{figure}

A perfect hash function maps a static set of n keys into a set of m integer numbers without collisions, where m is greater than or equal to n. If m is equal to n, the function is called minimal perfect hash function (MPHF) \cite{cmph} and is a bijection.
%As MPHF Algorithm we can mention CHD Algorithm\cite{botelho2007simple}, BDZ Algorithm\cite{botelho2008near}, BMZ Algorithm\cite{botelho2004new}, BRZ Algorithm\cite{botelho2004new}, CHM Algorithm\cite{czech1992optimal}, FCH Algorithm\cite{fox1992faster},etc.
One special class  of the MPHF is the order preserving MPHF (OPMPHF)~\cite{fox1989order}. As shown in Figure \ref{fig:opmhf_lay},  when  keys are defined in a certain order, OPMPHF returns the values in the exact order of the keys. In CMPH~\cite{cmph_concepts}, a perfect hash function is order preserving if the keys are arranged in some given order and preserve this order in the hash table.
In our system, we use the Monotone Minimal Perfect Hash Function (MMPHF)~\cite{belazzougui2009monotone} that will preserve the lexicographical ordering of keys.

%Each file's metadata in the index file has a fixed size of 24 bytes. If we can know in which position a file’s metadata was inserted into the index file, we could calculate its offset in the index file using equation \ref{equ:index_offset}. And after, we will just have to seek at this offset and to read the metadata without having to process the entire index file. This is where MMPHF will be useful, this hash function will help us to memorize the position in which a file's metadata was added in the index file.

\begin{figure}[h]
	\centering
		\includegraphics[width=\linewidth]{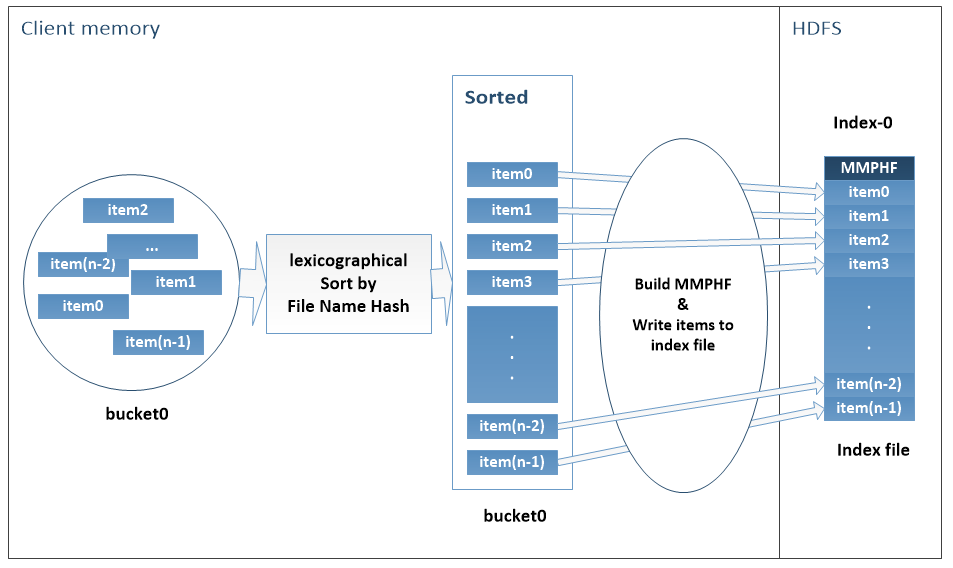}
	\caption{Build MMPHF \& Write index file}
	\label{fig:process_bucket}
\end{figure}

\begin{figure}[h]
	\centering
		\includegraphics[width=2in]{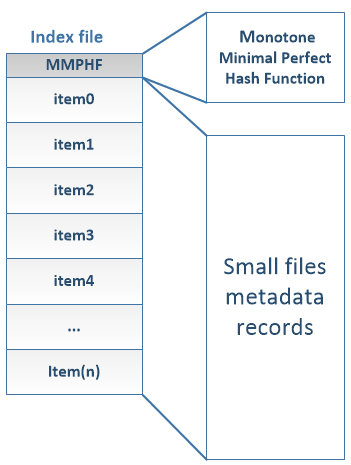}
	\caption{Index file format}
	\label{fig:index_file_lay}
\end{figure}

%As said above, the MMPHF needs the keys to be ordered in lexicographic order. 
At the end of the index file buckets creation process, we process each bucket as illustrated in Figure~\ref{fig:process_bucket}. We sort all metadata of each bucket according to file name hash values in lexicographic order. For each bucket, we collect all file name hash values representing the keys of the hash function, and we build the MMPHF. 
For a bucket and in its corresponding index file we write the MMPHF at the head of the index file, and the bucket records by maintaining the lexicographical ordering. When the contents of all the buckets are written, we delete the \_temporaryIndex file, which marks the end of the HPF file creation. Figure \ref{fig:index_file_lay} shows how each of the index files looks like, the header of the file contains the perfect hash function followed by files ’metadata. The entire creation process of HPF file is presented in Algorithm \ref{code:algo_creat}.

MMPHF does not consume a lot of memory, this function being calculated from the keys. In our design the keys are the hash code values of the file name.  It is interesting to note that MMPHF algorithms do not need to keep all  keys in memory. The majority of minimal perfect hash function requires less than 3 bits per key for their building. According to \cite{belazzougui2009monotone}, for a set $S$ of $n$ elements out of a universe of $2^{w}$ elements, $O(n log log w)$ bits are sufficient to hash monotonically with evaluation time $O(log w)$. We can get space $O(n log w)$ bits with $O(1)$ query time, this means it is possible to search a sorted table with $O(1)$ accesses to the table.

\begin{algorithm}[!ht]
\SetAlgoLined
 
$files$ = A set of small files\;
 
$buckets$ = EHF buckets\;
	
Create the data part and the temporaryIndex file\;

Create one bucket and add it to $buckets$\;

\tcc{Files merging}

\For {$f$ in $files$}{
  
Merge $f$ with part file\;

Get $f$ metadata\;

Append the metadata to the temporaryIndex file\;

Append the $f$ name to the names file\;

$bucket$ = Get from $buckets$ using EHF\;

Add $f$ metadata $bucket$\;

\If{$bucket$ is full}{%

$newBucket$ = Create new bucket and it index file\;

Redistribute data to $newBucket$ using EHF\;

Add the $newBucket$ to $buckets$\;

}

}

\tcc{Building index files}

\For {$b$ in $buckets$}{

Sort $b$ files metadata\;

Build the MMPHF\;

Write the MMPHF to the $b$ associated index file\;

Write all bucket Items to $b$ associated index file\;
	
}

\caption{HPF Creation Algorithm}
\label{code:algo_creat}
\end{algorithm}

\subsection{File Access \& Append }
\label{sec:FileAccess}

\begin{figure}[h]
	\centering
		\includegraphics[width=\linewidth]{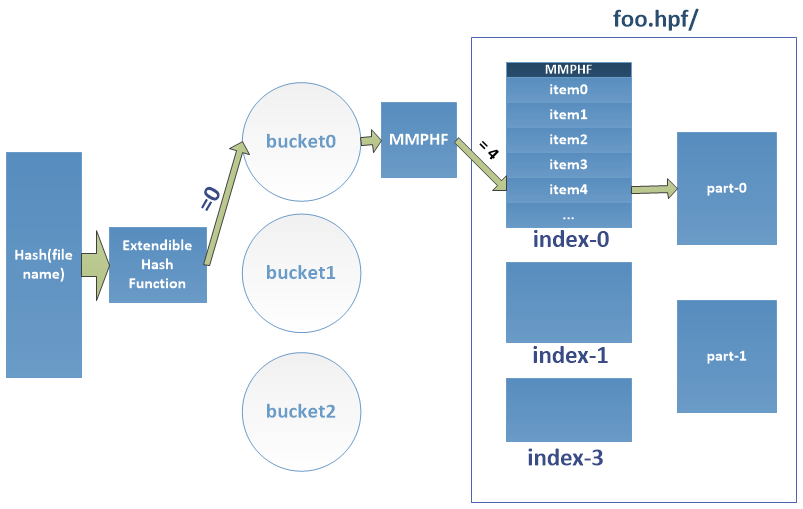}
	\caption{Files access in HPF file}
	\label{fig:hpf_access_file}
\end{figure}

The example of Figure \ref{fig:hpf_access_file} illustrates the 4 steps of the file's access in HPF:

(1) Firstly, we derive from the file name provided by the client the corresponding hash value.

(2) Secondly, EHF is used on the hash value to calculate the bucket position that also corresponds to the position of the index file that contains the file's metadata. If EHF return \textbf{i} it means the file metadata can be found in the index file named \textbf{index-i}.

(3) Thirdly, we use the bucket MMPHF to calculate the position of the file's metadata in the index file.
As shown in Table \ref{tab:IndexItemFieldsInformation} each file's metadata occupies 24Bytes in the index file, knowing metadata's position within the index file, we can calculate the exact offset of the file's metadata in the index file.
The computation of the offset is done using the Equation \ref{equ:index_offset}, to get the small file metadata we just have to read the index file from $offset$ to $offset +24 Bytes$.

\begin{equation}
		offset = \Upsilon + MMPHF(file\_key)*24 Bytes
\label{equ:index_offset}
\end{equation}
\textbf{Where:}
\begin{tabularx}{\linewidth}{>{$}r<{$} @{${}={}$} X}
\Upsilon & MMPHF size in index file.  \\
file\_key & file name hash code value.
\end{tabularx}
\newline\newline

(4) Finally, having the small file metadata (part file, offset in the part file, file size), with this information we access the part file and read of the file's content.

As we can notice to retrieve the metadata of a file, it is necessary to use two hash function functions. The first, the Extensible Hash Function helps to know in which index file is located a file’s metadata in $O(1)$ query time complexity. The second, the Monotone Minimal Perfect Hash Function, it helps to know at which offset of the index file a file’s metadata is located in $O(1)$ query time complexity.

\begin{figure}[h]
	\centering
		\includegraphics[width=\linewidth]{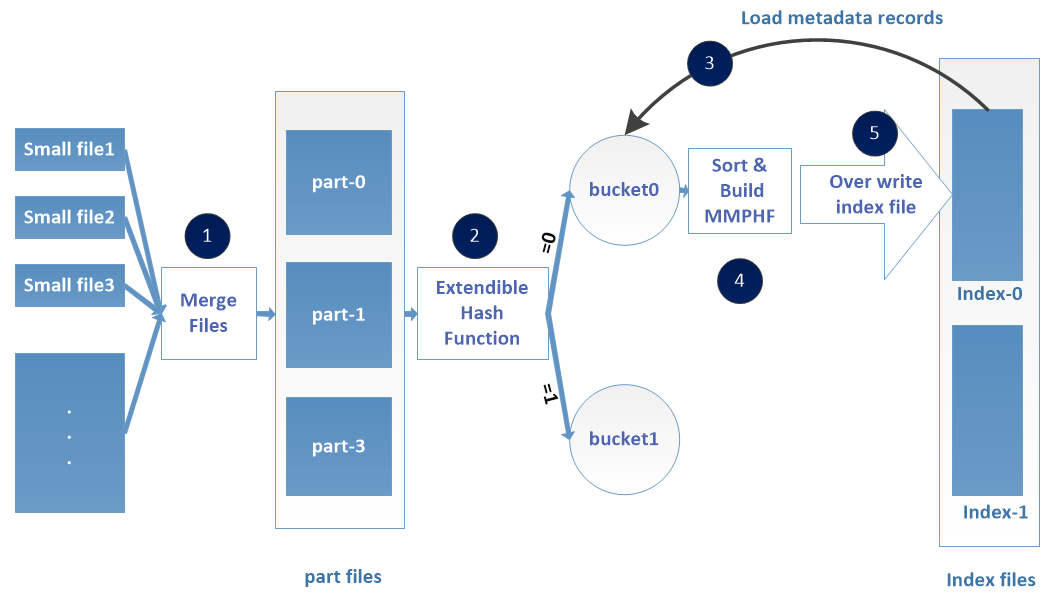}
	\caption{Files append in HPF file}
	\label{fig:hpf_add}
\end{figure}

The process of adding files to the HPF is almost identical to the creation process shown in Figure \ref{fig:hpf_creation}. 
Whenever the user wants to add more files, we operate like in Figure \ref{fig:hpf_add}:

(1) We recreate in HDFS the \_temporaryIndex file. 
We rebuild the merging threads to append the small files to be added to part-* files and append their names to the \_names files and their metadata to the \_temporaryIndex file. 

(2) We distribute the files’ metadata into buckets using the EHF used during the creation.

(3,4,5) The only difference from the creation process is before building the MMPHF, we have to reload into the buckets that have new records the content of their associated index file. For each of these buckets we sort, build again their MMPHF and overwrite the contents of their index file.

After having finished, we always don't forget to delete the \_temporaryIndex file.
HPF doesn’t force the user to sort and provide the files in a lexicographic order during its file creation or file appending as does the MapFile. That's why we have to rebuild and reorder the concerned index files whenever the user adds more files.

\section{Implementation Issues}
\label{sec:ImplementationOptimizations}

\subsection{Recovery on failure}
\label{sec:RecoveryOnFailure}

Building the index system at the client side after merging all files is not without risk, there are some advantages and some drawbacks in doing so. 
As advantages:
we are moving some of the processing out of the cluster,
the creation of the HPF file is faster because it avoids the massive network communications that could take place between the client and the servers of the cluster.
As drawback, the client is not reliable and can crash at any moment. 
This can interrupt the creation process or the files appending process of HPF.
During the files merging step, these files metadata are kept in buckets in the client memory and if the client crashes during the step, these files metadata will be lost and it will be impossible to restore the files appended to part files.
We have implemented a recovery on failure mechanism that allows us to avoid the metadata lost in case of client failure. You have probably noticed that each process of creation or adding files to the HPF file, we create a temporary index file (\_temporaryIndex). Whenever a file content is combined with the part-* file, we append the file's metadata with the temporaryIndex file before continuing the process. If the process finished without any problem, we delete the temporary index file, if a failure occurs at the client side during the process, this file is not deleted and contains all small files metadata that was combined with the part files.
So, the next time the client accesses the HPF file, we check if the archive contains the temporaryIndex file, if yes this means that a failure happened, so we rebuild our index files, delete this temporaryIndex file before giving access to the archive file.

%\subsection{Build the index system at the server side}
%\label{sec:PossibilityOfBuildingTheIndexSystemAtTheServerSide}
%
%As discussed in subsection \ref{sec:RecoveryOnFailure}, building the index system in client's memory is risky due to several failures that can occur, from where the need to set up a system of recovery in case of failure. 
The construction of our index system can be moved to the side of the NN in order to benefit from its High availability.
Once this option is indicated in the configuration files, when creating the HPF file or appending small files, at the first step which is the merging step, we append the files' contents to the part-* files, their names to the \_names files and the files' metadata to the temporaryIndex file only. After this first step, the client sends a request to the NN and the NN starts from the files' metadata present in the temporaryIndex file to build our index system by creating the buckets in its memory, building MMPHF and writing the index files to HDFS.
This high availability of the NN will help to face the problems that could happen in the case of client failure, but this time it is the resources (memory, processor) of the NN that will be used.

\subsection{Improving IOs Performance}
\label{sec:ImprovingIOsPerformance}

\subsubsection{Write performance}
\label{sec:WritePerformance}

\begin{figure}[h]
	\centering
		\includegraphics[width=\linewidth]{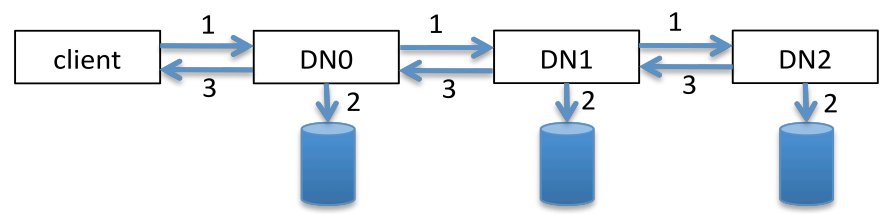}
	\caption{HDFS block Writing\&Replication process \cite{kuangappend}}
	\label{fig:hdfs_write}
\end{figure}

First, let's talk about how data is written in HDFS. 
To create a file or append data to a file on HDFS, the client interacts with the NN, the NN will provide to the client the addresses of the DNs on which the client starts writing the data. By default, HDFS performs three replicas for every data block on three different nodes; the client writes data on the first DN, the first DN writes on the second DN and the second DN writes on the third as shown in Figure \ref{fig:hdfs_write}, once the replicas created an acknowledgment is sent to the client before the client continues writing more data. Replication is done in series from one DN to another as can be seen in \cite{kuangappend}, not in parallel. During the data blocks writing to the DN storage space, blocks are written as normal files on the disk.

\begin{figure}[h]
	\centering
		\includegraphics[width=\linewidth]{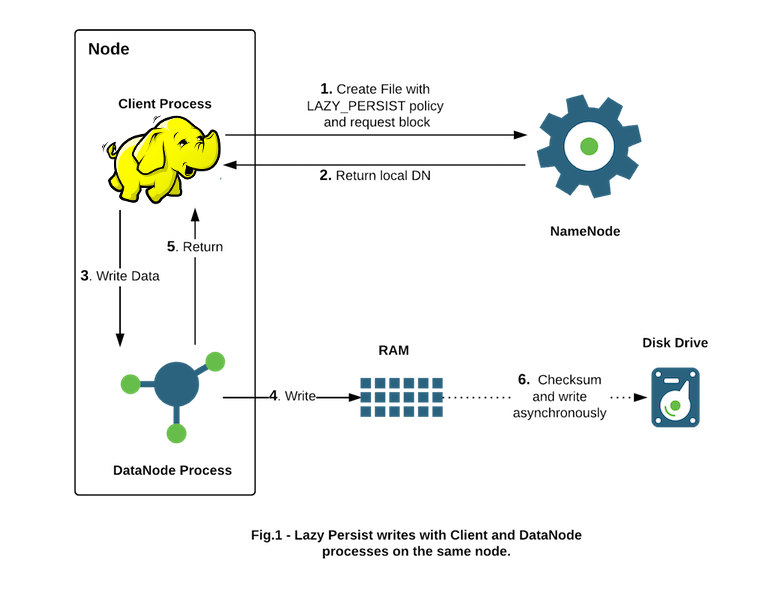}
	\caption{Lazy Persist Writes \cite{LazyP}}
	\label{fig:LazyPersistWrites}
\end{figure}

We have to notice that transferring data across the network and writing blocks on disk take a lot of time. For the data transfer, the problem can be the connection between the client and the first DN because it is often an external connection to the cluster and there is no guarantee on its performance. This external connection can be slower than the internal connection of the cluster (between the DNs and the NN) which often is more stable, reliable and high throughput. For slow disk writing, this is due to the fact that the majority of Hard Drives are mechanical.
To make the creation of HPF files faster, we used another data writing mode or \textbf{Storage Policy} proposed by Hadoop called the Lazy Persist write\cite{LazyP}.
In the Lazy Persist, data are written in each DN in an off-heap memory (See Figure \ref{fig:LazyPersistWrites}) located in the RAM. Writing in the off-heap memory is faster than writing on hard drive, it saves the client from waiting for the data to be written to the disk.
According to\cite{LazyP} The DataNodes will flush in-memory data to disk asynchronously thus removing expensive disk IO and checksum computations from the performance-sensitive IO path.
HDFS provides best-effort persistence guarantees for Lazy Persist Writes. Rare data loss is possible in the event of a node restart before replicas are persisted to disk.

We used LazyPersist storage policy in our approach to write the data in part-* files and speed up the creation process of our file. The limit of LazyPersist is that in version 2.9.1 of Hadoop that we used to perform our experiments, files created with the LazyPersist storage policy do not support the data-append functionality.
In order to maintain the HPF file appending functionality after its creation, after creating the HPF file, we reset the storage policy of all HPF part-* files to the default mode.
Our experiments show that the Lazy Persist Writing makes HPF file creation extremely faster compared to HAR file creation and MapFile creation.

\subsubsection{Read performances}
\label{sec:ReadPerformances}

Even if HPF file can know from which offset of the index file to where read file metadata, this still requires some IO operations. To completely eliminate the disc IOs operation during the metadata reading in the index files, and to increase the reading performance of the metadata within our index files, we use the Centralized Cache Management of HDFS. According to \cite{CentralizedCacheManagement}, the Centralized cache management in HDFS is an explicit caching mechanism that allows users to specify paths to be cached by HDFS. The NameNode will communicate with DataNodes that have the desired blocks on disk, and instruct them to cache the blocks in the off-heap caches.
This caching system allows us to tell the DN to maintain our index files in the off-heap memory, not on the disk. By doing so, we avoid the IO operation during metadata lookups and improve the time needed to restore the content of files without having to load the content of all the index files in client memory as MapFile or HAR file.

\subsection{File access performance analysis}
\label{sec:FileAccessAnalysis}

Let's call by $T_{Access\_metadata}$ the time needed to retrieve a file’s metadata from the index file(s) of an archive file and $T_{Access\_content}$, the time needed to restore the content of the file from the merged file. The access time of a file ($T_{Access}$) from the archive file is calculated by using Equation \ref{equ:accessTime}.
\begin{equation}
	T_{Access}=T_{Access\_metadata}+T_{Access\_content}
\label{equ:accessTime}
\end{equation}
If $T_{Access/HAR}$, $T_{Access/MapFile}$ and $T_{Access/HPF}$ are respectively the file’s access times in the HAR file, MapFile, and HPF files, we will have:
\setlength{\arraycolsep}{0.0em}
\begin{eqnarray}
T_{Access/HAR}=T_{Access\_metadata/HAR}\nonumber\\+T_{Access\_content/HAR}\\
\nonumber\\
T_{Access/MapFile}=T_{Access\_metadata/MapFile}\nonumber\\+T_{Access\_content/MapFile}\\
\nonumber\\
T_{Access/HPF}=T_{Access\_metadata/HPF}\nonumber\\+T_{Access\_content/HPF}
\end{eqnarray}
\setlength{\arraycolsep}{5pt}
Metadata access from the HAR file and the Map file require to read and process entirely all the index file(s) and since the HAR file’s index files are bigger and hold more information than MapFile’s index file, the metadata access from MapFile is faster than the metadata access from HAR file:
$T_{Acess\_metadata/MapFile}< T_{Acess\_metadata/HAR}$.
Metadata access from HPF index file(s) is direct and does not require reading and processing the entire index file. That's why the HPF file metadata access time is totally negligible compared to metadata access time in HAR file and MapFile as expressed by the following expression:
\setlength{\arraycolsep}{0.0em}
\begin{eqnarray}
		T_{Access\_metadata/HPF} \ll T_{Access\_metadata/MapFile}\nonumber\\< T_{Access\_metadata/HAR}\label{equ:metadatAccessTimeCompar}
\end{eqnarray}
\setlength{\arraycolsep}{5pt}
$T_{Access\_content}$, the time needed to restore a file’s content from the merged file can be different when the file is accessed from HAR file, MapFile or HPF files. This time is influenced by several factors like the cost of the seek operation, the use or not of compression algorithm to reduce the file size. 
If we assume that for the same file:
$T_{Acess\_content/HPF}= T_{Acess\_content/HAR}= T_{Acess\_content/MapFile}$
. According to equation \ref{equ:metadatAccessTimeCompar},
\begin{equation}
	T_{Acess/HPF}<T_{Acess/MapFile}< T_{Acess/HAR}
\label{equ:accessTimeComparAll}
\end{equation}
Equation \ref{equ:accessTimeComparAll} theoretically proves that file’s access in HPF file is faster than access to files in HAR and MapFile. This is also confirmed by our experiments presented in Subsection \ref{sec:TheAccessPerformanceWithoutCachingEffect}.

\section{Experimental evaluation}
\label{sec:ExperimentalEvaluation}

Our experiments evaluate the performance of HPF against existing solutions. We have implemented an open source prototype available on GitHub\footnote{The source code of Hadoop Perfect File is available at https://github.com/tchaye59/Hadoop-Perfect-File}. We performed some experiments and compared HPF performance against the HAR file, the MapFile and also compared to the native HDFS. We took into account criteria such as access time, creation time, memory usage of NameNode.

%\subsection{Experiment Environment}
%\label{sec:ExperimentEnvironment}

Our experimental test platform is built on a cluster of 6 nodes.
The node that acts as NameNode is a server of two CPU cores of 2.13GHz each, 8GB for RAM and 500GB Hard Disk.  
The other 5 nodes, act as DataNodes are also server's of two CPU cores of 2.13GHz each, 8GB for RAM and 500GB Hard Disk. 
For all nodes, the operating system is Ubuntu GNU/Linux 4.10.0-28-generic x86\textunderscore64.
The client is a Lenovo ThinkPad E560 Laptop, is running on Microsoft Windows 10 Pro operating system and has 16GB of RAM,1TB of disk, Processor Intel(R) Core(TM) i7-6500U CPU @ 2.50GHz, 2601 Mhz, 2 Core(s), 4 Logical Processor(s).
Hadoop version is 2.9.1 and JDK version is jdk1.8.0\textunderscore102.
The number of replicas is set to 3 and HDFS block size is let to it default value 128 MB during the tests. 
For the test purpose we use logs’ text files from different servers. 
We use multiple files sets containing 100000, 200000, 300000 and 400000 files, the total size is respectively 1.44 GB, 2.37 GB, 3.30 GB and 4.23 GB.
File sizes range from 1 KB to 10 MB.

\subsection{Experiments}
\label{sec:Experiments}

The first category of tests aims to evaluate the small files access performance in HPF, HAR, MapFile and also their performance when accessed directly from HDFS.
The second category of tests aims to evaluate the performances related to the creation of HAR file, MapFile and HPF file like the creation time cost, the DataNodes disk space usage, the NameNode memory usage.
In our experiments, we set the maximum capacity of HPF index file's bucket to 200000 records. We increase our part files block size to 512MB to make each block contain more files.
This mechanism makes the HPF efficiently manage a very large number of files by requiring little amount of metadata.
We run all our experiments several times in order to eliminate errors that may be due to network congestion or other errors.

\subsection{The access performance of small files}
\label{sec:TheAccessPerformanceOfSmallFiles}

We evaluate the access performance of small files by randomly accessing 100 files from HDFS and in HPF file, MapFile, HAR file.
As mentioned in subsection \ref{sec:FileAccessFromArchiveFileWithTheCachingEffect}, the MapFile and HAR file cache in the client's memory all the metadata of their files during the first access in order to improve their access performance. 
To see the real performances of HAR and MapFile compared to HPF file, we firstly evaluated the access performance without caching effect and after the access performance with caching effect.

\subsubsection{The access performance without caching effect}
\label{sec:TheAccessPerformanceWithoutCachingEffect}

\begin{figure}[htbp]
	\begin{subfigure}{.5\linewidth}
		\centering
		\includegraphics[width=\linewidth]{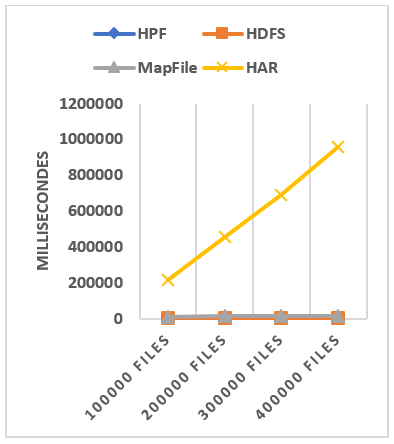}
		\caption{ }
		\label{fig:exp_access_time_nocache1}
	\end{subfigure}%
	\begin{subfigure}{.5\linewidth}
		\centering
		\includegraphics[width=\linewidth]{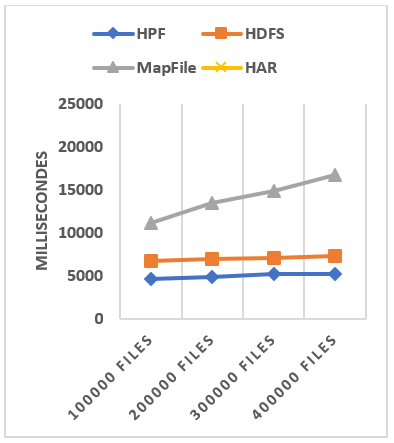}
		\caption{ }
		\label{fig:exp_access_time_nocache2}
	\end{subfigure}
	\caption{The access performance without caching}
	\label{fig:exp_access_time_nocache}
\end{figure}

To disable the caching effect of HAR file and MapFile, we create a new access object at each file's access.  
Table \ref{tab:access_time_no_cache} collects the file access time for each dataset and technique.
The value in parenthesis in the Table represent the percentage at which our solution is faster than the used technique. It is calculated by doing: $100*\frac{X - Y}{Y}$.
Where $X$ represents the used technique access time  and $Y$ HPF access time.
The access performance comparison is illustrated in Figure \ref{fig:exp_access_time_nocache}.

\begin{table}[htbp]
\centering
\resizebox{\linewidth}{!}{%
\begin{tabular}{|l|c|c|c|c|} 
\hline
Fs/Set~ & 100000 files                                              & 200000 files                                              & 300000 files                                               & 400000 files                                               \\ 
\hline
HPF     & 4680                                                      & 4844                                                      & 5220                                                       & 5253                                                       \\ 
\hline
HDFS    & \begin{tabular}[c]{@{}c@{}}6712\\(43\%) \end{tabular}     & \begin{tabular}[c]{@{}c@{}}6966\\(44\%) \end{tabular}     & \begin{tabular}[c]{@{}c@{}}7068\\(35\%) \end{tabular}      & \begin{tabular}[c]{@{}c@{}}7302\\(39\%)\end{tabular}       \\ 
\hline
MapFile & \begin{tabular}[c]{@{}c@{}}11081\\(137\%) \end{tabular}   & \begin{tabular}[c]{@{}c@{}}13469\\(178\%) \end{tabular}   & \begin{tabular}[c]{@{}c@{}}14825\\(184\%) \end{tabular}    & \begin{tabular}[c]{@{}c@{}}16668\\(217\%)\end{tabular}     \\ 
\hline
HAR     & \begin{tabular}[c]{@{}c@{}}217448\\(4546\%) \end{tabular} & \begin{tabular}[c]{@{}c@{}}457647\\(9348\%) \end{tabular} & \begin{tabular}[c]{@{}c@{}}691009\\(13138\%) \end{tabular} & \begin{tabular}[c]{@{}c@{}}958316\\(18143\%)\end{tabular}  \\
\hline
\end{tabular}
}
\caption{Access times without caching(milliseconds)}
\label{tab:access_time_no_cache}
\end{table}

As shown in Figure \ref{fig:exp_access_time_nocache1}, the performance of the HAR without the use of caching is really bad and evolves linearly when the dataset contains more files. In Figure \ref{fig:exp_access_time_nocache2}, we clearly can see the performances of the other techniques.
HPF is faster than the original HDFS, MapFile, and HAR.
By analyzing the data in Table \ref{tab:access_time_no_cache} and taking the average percentage of each technique, it can be seen that HPF can significantly outperform other file systems.
The small files' access in HPF can be more than 40\% faster than the access in the original HDFS, 179\% faster than the access in MapFile and 11294\% faster than the access in HAR file without considering the caching effect.
These performances of HPF are due to the facts that:
\begin{itemize}
	\item To get file metadata without the caching effect, at every file access MapFile and HAR file read and process their whole index file(s) so need more time to get the metadata which is not the case of the HPF file.
	
	\item MapFile needs two decompression level  which makes it slower compare to HPF where only one decompression level is required.

	\item HDFS keeps its metadata in memory of NN and HPF keeps its index files in memory of DNs using Hadoop centralized cache management system.
	
	\item In the case where the file is stored directly on HDFS, the communication between the client and the NN in order to get the file metadata is done by using the RPC(Remote Procedure Call) protocol. This protocol is built on top of the Sockets protocol and is slower than Sockets.
	In the case of HPF, files communication is done between the client and the DN in order to get the file metadata by using sockets which are faster than RPC calls. To read and write files on HDFS, Hadoop uses sockets, HPF just performs a file read operation on a portion of the index file to read file's metadata.
					
\end{itemize}

 \subsubsection{The access performance with caching effect}
 \label{sec:TheAccessPerformanceWithCachingEffect}

\begin{figure}[h]
	\centering
		\includegraphics[width=\linewidth]{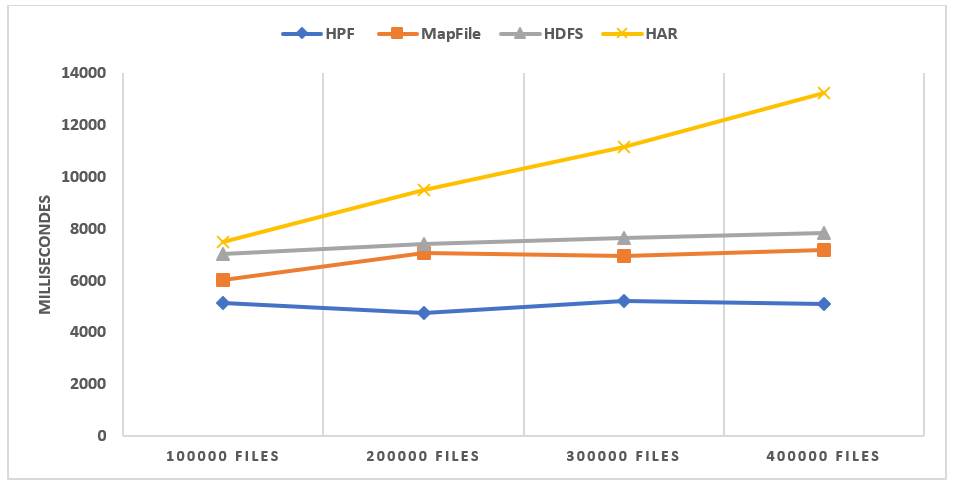}
	\caption{The access performance with caching}
	\label{fig:exp_access_time_cache}
\end{figure}

We rerun the experiments taking into account the caching effect, and we collected the access times of each technique and dataset in Table \ref{tab:exp_access_time_cache}.
The access performance comparison is illustrated in Figure \ref{fig:exp_access_time_cache}.
In this experiment, since only HAR and MapFile cache their metadata in client memory, we notice a big difference in their performances which are greatly improved.
Despite this improvement HPF still the fastest,  MapFile comes in second position and is slightly faster than the access from the original HDFS due to the fact that to access from HDFS still need to make two requests (one to get the file metadata and one for the content of the file) while in the case of MapFile, all metadata are loaded in memory of the client. 
Figure \ref{fig:exp_access_time_cache} results show us that HPF is faster than MapFile in term of random access. This is because MapFile is optimized to access files in the lexicographical order in which the files were added, not randomly.

\begin{table}[htbp]
\centering
\resizebox{\linewidth}{!}{%
\begin{tabular}{|l|c|c|c|c|} 
\hline
Fs/Set~ & 100000 files                                          & 200000 files                                           & 300000 files                                            & 400000 files                                            \\ 
\hline
HPF     & 5135                                                  & 4749                                                   & 5212                                                    & 5086                                                    \\ 
\hline
MapFile & \begin{tabular}[c]{@{}c@{}}6025\\(17\%) \end{tabular} & \begin{tabular}[c]{@{}c@{}}7058\\(49\%) \end{tabular}  & \begin{tabular}[c]{@{}c@{}}6932\\(33\%) \end{tabular}   & \begin{tabular}[c]{@{}c@{}}7173\\(41\%)\end{tabular}    \\ 
\hline
HDFS    & \begin{tabular}[c]{@{}c@{}}7030\\(37\%) \end{tabular} & \begin{tabular}[c]{@{}c@{}}7404\\(56\%) \end{tabular}  & \begin{tabular}[c]{@{}c@{}}7637\\(47\%) \end{tabular}   & \begin{tabular}[c]{@{}c@{}}7820\\(54\%)\end{tabular}    \\ 
\hline
HAR     & \begin{tabular}[c]{@{}c@{}}7483\\(46\%) \end{tabular} & \begin{tabular}[c]{@{}c@{}}9488\\(100\%) \end{tabular} & \begin{tabular}[c]{@{}c@{}}11144\\(114\%) \end{tabular} & \begin{tabular}[c]{@{}c@{}}13232\\(160\%)\end{tabular}  \\
\hline
\end{tabular}
}
\caption{Access times with caching(milliseconds)}
\label{tab:exp_access_time_cache}
\end{table}

Table \ref{tab:exp_access_time_cache} data shows that HPF still can significantly outperform other file systems.
The small file's access in HPF can be more than 48\% faster than the access in the original HDFS, 35\% faster than the access in MapFile and 105\% faster than the access in HAR file if we consider the caching effect.

\subsection{Archive file’s creation efficiency}
\label{sec:ArchiveFileSCreationEfficiency}

\begin{figure}[htbp]
	\begin{subfigure}{.5\linewidth}
		\centering
		\includegraphics[width=\linewidth]{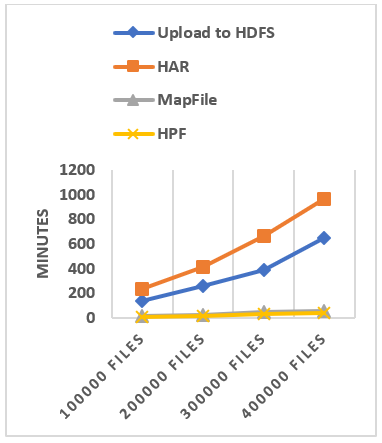}
		\caption{ }
		\label{fig:exp_creating_time1}
	\end{subfigure}%
	\begin{subfigure}{.5\linewidth}
		\centering
		\includegraphics[width=\linewidth]{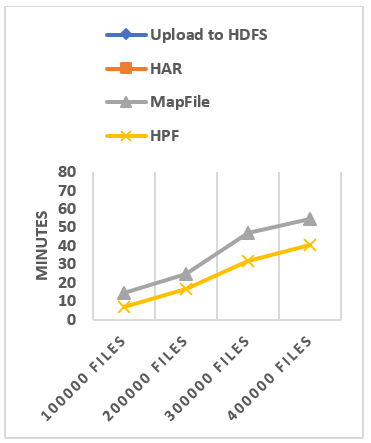}
		\caption{ }
		\label{fig:exp_creating_time2}
	\end{subfigure}
	\caption{The performance of creating a new archive file}
	\label{fig:exp_creating_time}
\end{figure}

We have evaluated the time spent in creating HAR file, MapFile and HPF file for each dataset and the results are shown in Figure \ref{fig:exp_creating_time}.
It can be seen in Figure \ref{fig:exp_creating_time2} that HPF have the fastest creation time for each dataset and is followed by the MapFile. For Figure \ref{fig:exp_creating_time1} it can be seen that HAR is the slowest and takes a lot of time compared to other solutions. We notice that the creation time of the MapFile and HPF seems completely negligible compared to the time needed to create the HAR file, this is due to many factors.
Despite the fact that the HAR uses MapReduce and handles small files in parallel, HPF and MapFile creation are faster because they do not need to upload all small files to HDFS before their creation. For the HAR, we have to upload the dataset to HDFS and then launch MapReduce Job to create HAR file. We calculate the creation time of the HAR file by adding the dataset upload time with the time the MapReduce Job takes to generate the HAR file.
The second reason for the slowness of HAR file is also the disk and network IO operations. With the MapFile and HPF file, the data is compressed at the client side before being sent through the network. This allows them to reduce the amount of data to be sent across the network and the amount of data to be written to the DNs' discs. MapFile and HPF file have better use of the bandwidth compare with the HAR which can lead easily to network congestion.
The third reason for the slowness of the HAR is due to it MapReduce job.
The HAR MapReduce job takes too much time to prepare its map tasks splits because it needs to get each file metadata from NameNode, which takes a lot of time with a large number of files.

The results of Figure \ref{fig:exp_creating_time2} shows that our approach is slightly faster than MapFile, the reason can be because 
the file merging process is done in parallel. We use of the LazyPersist writes and also because, in our prototype of HPF file, the compression is applied only to each file (at record level) while MapFile takes more time by doing the compression at the record level and at block level. And also the MapFile creation doesn’t support parallelism like HPF and HAR file.
We use in our prototype the LZ4\cite{lz4} compression algorithm. LZ4 is extremely fast and able to achieve a compression rate higher than 500 MB/s per core and a decompression rate of multiple GB/s per core.

\subsection{NameNode’s Memory usage}
\label{sec:NameNodeSMemoryUsage}

\begin{figure}[htbp]
	\begin{subfigure}{.5\linewidth}
		\centering
		\includegraphics[width=\linewidth]{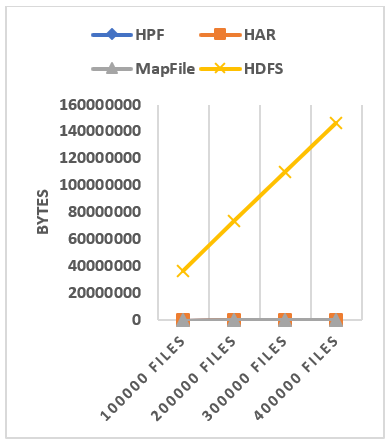}
		\caption{ }
		\label{fig:exp_nn_memory_use1}
	\end{subfigure}%
	\begin{subfigure}{.5\linewidth}
		\centering
		\includegraphics[width=\linewidth]{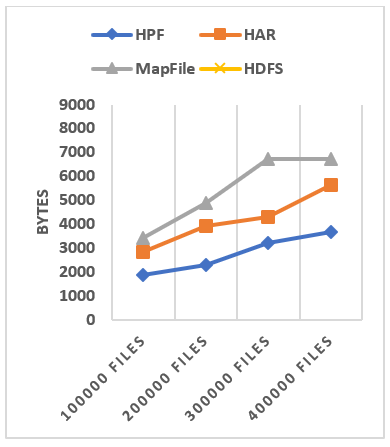}
		\caption{ }
		\label{fig:exp_nn_memory_use2}
	\end{subfigure}
	\caption{NameNode Memory usage}
	\label{fig:exp_nn_memory_use}
\end{figure}

NameNode memory used by each dataset's metadata when it is directly stored in HDFS and when it is stored using the HAR, MapFile and HPF systems is also evaluated. The results are shown in Figure \ref{fig:exp_nn_memory_use}.
Figure \ref{fig:exp_nn_memory_use1} shows that HAR, MapFile, and HPF file are more efficient for storing small files than HDFS. They consume less memory within the NameNode than the native HDFS.
From Figure \ref{fig:exp_nn_memory_use2} we can see that HAR file and the HPF file use slightly less metadata than MapFile because MapFile uses the default HDFS block size (128MB) while the HAR files and our approach uses a larger block of 512 MB. This makes HAR files and HPF files using fewer blocks for the same dataset than MapFile.

\subsection{Archive files sizes after creation}
\label{sec:ArchiveFileSSizeAfterCreation}

\begin{figure}[h]
	\centering
		\includegraphics[width=\linewidth]{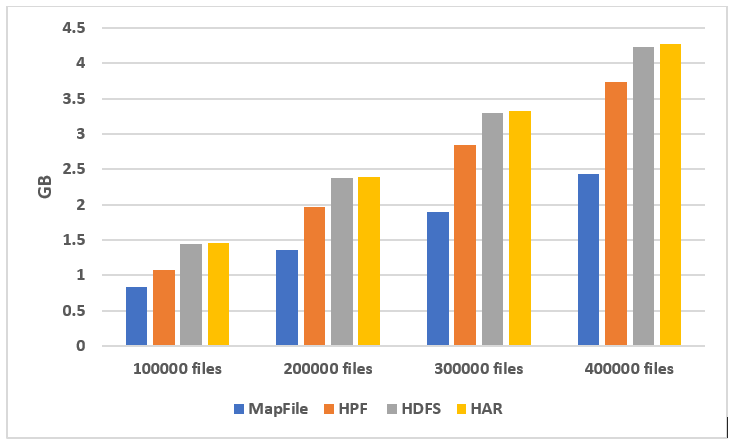}
	\caption{File size}
	\label{fig:exp_files_sizes}
\end{figure}

We collected the size of each archive file after its creation which gives us an idea of the disk storage space used from the DNs.
In Figure \ref{fig:exp_files_sizes}, we have the size of each archive file storing 100000, 200000, 300000, and 400000 small files.
Since the HAR does not use any compression algorithms, the size of the HAR file for each dataset is almost equal to the size of the dataset when stored directly on HDFS.

The size of the HPF and MapFile files are reduced due to compression, our analysis shows that in our experiment and for all dataset, MapFile saves more than 42\% of disc space and HPF save more than 11\% of disc space.
This difference is because with MapFile the compression is done at two levels whereas in our HPF prototype compression is only performed at the record level. MapFile and HPF files optimally use the storage space than the HAR files.

\section{Conclusion and future work}
\label{sec:ConclusionAndFutureWork}
 HDFS has only been thought and built to provide maximum performance with large files. 
Implementing a solution to solve the small file problem in HDFS requires taking into account the memory overflow of the NN and providing fast access to small files. 
The previous work mainly focused on building archive systems where they merge small files into large files and use these small files metadata to build index files before storing on HDFS. 
This approach effectively reduces the metadata load in NN's memory but leads to weak file-access performance. 
In this paper, we have presented a new type of index-based archive file called HPF optimize to provide fast access to data from HDFS and support the appending of more files after the creation of HPF file. 
To ensure the seek operation efficient performance, HPF uses an extensible hash function to distribute its metadata in several index files before building MMPHF for each index file. 
HPF access file's metadata in a different way, the metadata of each file in HPF has a fixed size (24Bytes in our prototype) with it monotone minimal perfect hash table, HPF only read the part of the index file that contains the searched file metadata instead of the entire index files. In order to totally eliminate the surplus of IOs operations during metadata lookup, HPF uses the centralized cache management of Hadoop to keep its index files in memory of DNs. Each file access in HPF needs only one disk operation to read the file content. 
Our experiments show that HPF can significantly outperform other file systems like HAR, MapFile and the original HDFS. 
For future works, we planned the following directions: 
\begin{enumerate} 
\item We will study the memory size consumed by our two hash functions information (EHF and MMPHF) in client's memory. 
\item We will study the possibility of implementing compression at the block level as with MapFile. 
\item We will study the possibility of implementing files deletion functionality. 
\end{enumerate}

\ifCLASSOPTIONcompsoc
  \section*{Acknowledgments}
\else
  \section*{Acknowledgment}
\fi
The authors thank the anonymous reviewers for their insightful suggestions. This work is supported by the National Nature Science Foundation of China (Grant No. 61602037).

\ifCLASSOPTIONcaptionsoff
  \newpage
\fi

\Urlmuskip=0mu plus 1mu\relax  
\bibliographystyle{IEEEtran}
\input{references.bbl}

\end{document}

%% file: references.bbl

%% file: main_full.bbl
\begin{thebibliography}{10}
\providecommand{\url}[1]{#1}
\csname url@samestyle\endcsname
\providecommand{\newblock}{\relax}
\providecommand{\bibinfo}[2]{#2}
\providecommand{\BIBentrySTDinterwordspacing}{\spaceskip=0pt\relax}
\providecommand{\BIBentryALTinterwordstretchfactor}{4}
\providecommand{\BIBentryALTinterwordspacing}{\spaceskip=\fontdimen2\font plus
\BIBentryALTinterwordstretchfactor\fontdimen3\font minus
  \fontdimen4\font\relax}
\providecommand{\BIBforeignlanguage}[2]{{%
\expandafter\ifx\csname l@#1\endcsname\relax
\typeout{** WARNING: IEEEtran.bst: No hyphenation pattern has been}%
\typeout{** loaded for the language `#1'. Using the pattern for}%
\typeout{** the default language instead.}%
\else
\language=\csname l@#1\endcsname
\fi
#2}}
\providecommand{\BIBdecl}{\relax}
\BIBdecl

\bibitem{shvachko2010hadoop}
K.~Shvachko, H.~Kuang, S.~Radia, and R.~Chansler, ``The hadoop distributed file
  system,'' in \emph{Mass storage systems and technologies (MSST), 2010 IEEE
  26th symposium on}.\hskip 1em plus 0.5em minus 0.4em\relax Ieee, 2010, pp.
  1--10.

\bibitem{ghemawat2003google}
S.~Ghemawat, H.~Gobioff, and S.-T. Leung, \emph{The Google file system}.\hskip
  1em plus 0.5em minus 0.4em\relax ACM, 2003, vol.~37, no.~5.

\bibitem{dean2008mapreduce}
J.~Dean and S.~Ghemawat, ``Mapreduce: simplified data processing on large
  clusters,'' \emph{Communications of the ACM}, vol.~51, no.~1, pp. 107--113,
  2008.

\bibitem{harSite}
Hadoop archives guide.
  \url{https://hadoop.apache.org/docs/r2.7.5/hadoop-archives/HadoopArchives.html}.
  Accessed: 2018-12-15.

\bibitem{White15}
\BIBentryALTinterwordspacing
T.~White, \emph{Hadoop: The Definitive Guide}, 4th~ed.\hskip 1em plus 0.5em
  minus 0.4em\relax Beijing: O'Reilly, 2015. [Online]. Available:
  \url{https://www.safaribooksonline.com/library/view/hadoop-the-definitive/9781491901687/}
\BIBentrySTDinterwordspacing

\bibitem{zheng2017method}
T.~Zheng, W.~Guo, and G.~Fan, ``A method to improve the performance for storing
  massive small files in hadoop,'' in \emph{7th International Conference on
  Computer Engineering and Networks}, 2017.

\bibitem{nhar}
C.~Vorapongkitipun and N.~Nupairoj, ``Improving performance of small-file
  accessing in hadoop,'' in \emph{Computer Science and Software Engineering
  (JCSSE), 2014 11th International Joint Conference on}.\hskip 1em plus 0.5em
  minus 0.4em\relax IEEE, 2014, pp. 200--205.

\bibitem{bok2017efficient}
K.~Bok, H.~Oh, J.~Lim, Y.~Pae, H.~Choi, B.~Lee, and J.~Yoo, ``An efficient
  distributed caching for accessing small files in hdfs,'' \emph{Cluster
  Computing}, vol.~20, no.~4, pp. 3579--3592, 2017.

\bibitem{dong2010novel}
B.~Dong, J.~Qiu, Q.~Zheng, X.~Zhong, J.~Li, and Y.~Li, ``A novel approach to
  improving the efficiency of storing and accessing small files on hadoop: a
  case study by powerpoint files,'' in \emph{Services Computing (SCC), 2010
  IEEE International Conference on}.\hskip 1em plus 0.5em minus 0.4em\relax
  IEEE, 2010, pp. 65--72.

\bibitem{sheoran2017optimized}
S.~Sheoran, D.~Sethia, and H.~Saran, ``Optimized mapfile based storage of small
  files in hadoop,'' in \emph{Cluster, Cloud and Grid Computing (CCGRID), 2017
  17th IEEE/ACM International Symposium on}.\hskip 1em plus 0.5em minus
  0.4em\relax IEEE, 2017, pp. 906--912.

\bibitem{meng2016novel}
B.~Meng, W.-b. Guo, G.-s. Fan, and N.-w. Qian, ``A novel approach for efficient
  accessing of small files in hdfs: Tlb-mapfile,'' in \emph{Software
  Engineering, Artificial Intelligence, Networking and Parallel/Distributed
  Computing (SNPD), 2016 17th IEEE/ACIS International Conference on}.\hskip 1em
  plus 0.5em minus 0.4em\relax IEEE, 2016, pp. 681--686.

\bibitem{peng2018hadoop}
J.-f. Peng, W.-g. Wei, H.-m. Zhao, Q.-y. Dai, G.-y. Xie, J.~Cai, and K.-j. He,
  ``Hadoop massive small file merging technology based on visiting hot-spot and
  associated file optimization,'' in \emph{International Conference on Brain
  Inspired Cognitive Systems}.\hskip 1em plus 0.5em minus 0.4em\relax Springer,
  2018, pp. 517--524.

\bibitem{huo2016sfs}
Y.~Huo, Z.~Wang, X.~Zeng, Y.~Yang, W.~Li, and C.~Zhong, ``Sfs: A massive small
  file processing middleware in hadoop,'' in \emph{Network Operations and
  Management Symposium (APNOMS), 2016 18th Asia-Pacific}.\hskip 1em plus 0.5em
  minus 0.4em\relax IEEE, 2016, pp. 1--4.

\bibitem{EasyChair:773}
W.~Tao, Y.~Zhai, and J.~Tchaye-Kondi, ``Lhf: A new archive based approach to
  accelerate massive small files access performance in hdfs,'' in \emph{in
  Proceeding of The Fifth IEEE International Conference On Big Data Service And
  Applications}, 2019.

\bibitem{jing2018optimized}
W.~Jing, D.~Tong, G.~Chen, C.~Zhao, and L.~Zhu, ``An optimized method of hdfs
  for massive small files storage,'' \emph{Computer Science and Information
  Systems}, vol.~15, no.~3, pp. 533--548, 2018.

\bibitem{he2016optimization}
H.~He, Z.~Du, W.~Zhang, and A.~Chen, ``Optimization strategy of hadoop small
  file storage for big data in healthcare,'' \emph{The Journal of
  Supercomputing}, vol.~72, no.~10, pp. 3696--3707, 2016.

\bibitem{cai2018optimization}
X.~Cai, C.~Chen, and Y.~Liang, ``An optimization strategy of massive small
  files storage based on hdfs,'' in \emph{2018 Joint International Advanced
  Engineering and Technology Research Conference (JIAET 2018)}.\hskip 1em plus
  0.5em minus 0.4em\relax Atlantis Press, 2018.

\bibitem{tfs}
Tfs. \url{https://github.com/alibaba/tfs}. Accessed: 2018-12-15.

\bibitem{beaver2010finding}
D.~Beaver, S.~Kumar, H.~C. Li, J.~Sobel, P.~Vajgel \emph{et~al.}, ``Finding a
  needle in haystack: Facebook's photo storage.'' in \emph{OSDI}, vol.~10, no.
  2010, 2010, pp. 1--8.

\bibitem{lakshman2010cassandra}
A.~Lakshman and P.~Malik, ``Cassandra: a decentralized structured storage
  system,'' \emph{ACM SIGOPS Operating Systems Review}, vol.~44, no.~2, pp.
  35--40, 2010.

\bibitem{fu2015performance}
S.~Fu, L.~He, C.~Huang, X.~Liao, and K.~Li, ``Performance optimization for
  managing massive numbers of small files in distributed file systems,''
  \emph{IEEE Transactions on Parallel and Distributed Systems}, vol.~26,
  no.~12, pp. 3433--3448, 2015.

\bibitem{phakade2014innovative}
P.~Phakade and S.~Raut, ``An innovative strategy for improved processing of
  small files in hadoop,'' \emph{International Journal of Application of
  Innovation in Enginnering \& Management (IJAIEM)}, pp. 278--280, 2014.

\bibitem{choi2018improved}
C.~Choi, C.~Choi, J.~Choi, and P.~Kim, ``Improved performance optimization for
  massive small files in cloud computing environment,'' \emph{Annals of
  Operations Research}, vol. 265, no.~2, pp. 305--317, 2018.

\bibitem{niazi2018size}
S.~Niazi, M.~Ronstr{\"o}m, S.~Haridi, and J.~Dowling, ``Size matters: Improving
  the performance of small files in hadoop,'' in \emph{Proceedings of the 19th
  International Middleware Conference}.\hskip 1em plus 0.5em minus 0.4em\relax
  ACM, 2018, pp. 26--39.

\bibitem{kim2017improving}
H.~Kim and H.~Yeom, ``Improving small file i/o performance for massive digital
  archives,'' in \emph{e-Science (e-Science), 2017 IEEE 13th International
  Conference on}.\hskip 1em plus 0.5em minus 0.4em\relax IEEE, 2017, pp.
  256--265.

\bibitem{niazi2017hopsfs}
S.~Niazi, M.~Ismail, S.~Haridi, J.~Dowling, S.~Grohsschmiedt, and
  M.~Ronstr{\"o}m, ``Hopsfs: Scaling hierarchical file system metadata using
  newsql databases.'' in \emph{FAST}, 2017, pp. 89--104.

\bibitem{shvachko2007name}
K.~Shvachko, ``Name-node memory size estimates and optimization proposal,''
  \emph{Apache Hadoop Common Issues, HADOOP-1687}, 2007.

\bibitem{dan1990approximate}
A.~Dan and D.~Towsley, \emph{An approximate analysis of the LRU and FIFO buffer
  replacement schemes}.\hskip 1em plus 0.5em minus 0.4em\relax ACM, 1990,
  vol.~18, no.~1.

\bibitem{CentralizedCacheManagement}
Centralized cache management in hdfs.
  \url{http://hadoop.apache.org/docs/stable/hadoop-project-dist/hadoop-hdfs/CentralizedCacheManagement.html}.
  Accessed: 2018-12-15.

\bibitem{fagin1979extendible}
R.~Fagin, J.~Nievergelt, N.~Pippenger, and H.~R. Strong, ``Extendible
  hashing---a fast access method for dynamic files,'' \emph{ACM Transactions on
  Database Systems (TODS)}, vol.~4, no.~3, pp. 315--344, 1979.

\bibitem{zhang2009extendible}
D.~Zhang, Y.~Manolopoulos, Y.~Theodoridis, and V.~J. Tsotras, ``Extendible
  hashing,'' in \emph{Encyclopedia of Database Systems}.\hskip 1em plus 0.5em
  minus 0.4em\relax Springer, 2009, pp. 1093--1095.

\bibitem{fox1989order}
E.~A. Fox, Q.~F. Chen, A.~M. Daoud, and L.~S. Heath, ``Order preserving minimal
  perfect hash functions and information retrieval,'' in \emph{Proceedings of
  the 13th annual international ACM SIGIR conference on Research and
  development in information retrieval}.\hskip 1em plus 0.5em minus 0.4em\relax
  ACM, 1989, pp. 279--311.

\bibitem{peb_eh}
Extendible hashing data structure.
  \url{https://xlinux.nist.gov/dads/HTML/extendibleHashing.html}. Accessed:
  2018-12-15.

\bibitem{Fagin:1979:EHF:320083.320092}
\BIBentryALTinterwordspacing
R.~Fagin, J.~Nievergelt, N.~Pippenger, and H.~R. Strong, ``Extendible
  hashing\&mdash;a fast access method for dynamic files,'' \emph{ACM Trans.
  Database Syst.}, vol.~4, no.~3, pp. 315--344, Sep. 1979. [Online]. Available:
  \url{http://doi.acm.org/10.1145/320083.320092}
\BIBentrySTDinterwordspacing

\bibitem{ExtendedAttributes}
Extended attributes in hdfs.
  \url{https://hadoop.apache.org/docs/r2.9.2/hadoop-project-dist/hadoop-hdfs/ExtendedAttributes.html}.
  Accessed: 2018-12-15.

\bibitem{cmph}
Cmph - c minimal perfect hashing library. \url{http://cmph.sourceforge.net/}.
  Accessed: 2018-12-15.

\bibitem{cmph_concepts}
Minimal perfect hash functions - introduction.
  \url{http://cmph.sourceforge.net/concepts.html}. Accessed: 2018-12-15.

\bibitem{belazzougui2009monotone}
D.~Belazzougui, P.~Boldi, R.~Pagh, and S.~Vigna, ``Monotone minimal perfect
  hashing: searching a sorted table with o (1) accesses,'' in \emph{Proceedings
  of the twentieth annual ACM-SIAM symposium on Discrete algorithms}.\hskip 1em
  plus 0.5em minus 0.4em\relax SIAM, 2009, pp. 785--794.

\bibitem{kuangappend}
H.~Kuang, K.~Shvachko, N.~Sze, S.~Radia, and R.~Chansler, ``Append/hflush/read
  design.''

\bibitem{LazyP}
Memorystorage.
  \url{http://hadoop.apache.org/docs/stable/hadoop-project-dist/hadoop-hdfs/MemoryStorage.html}.
  Accessed: 2018-12-15.

\bibitem{lz4}
lz4. \url{https://github.com/lz4/lz4}. Accessed: 2018-12-15.

\end{thebibliography}
